\documentclass[
 reprint,
 showpacs, preprintnumbers,
 amsmath, amssymb,
 aps,
 prl,
 floatfix,
 superscriptaddress
]{revtex4-2}

\usepackage{amsmath}
\usepackage{graphicx}% Include figure files
\usepackage{dcolumn}% Align table columns on decimal point
\usepackage{bm}% bold math
\usepackage{bbm}
\usepackage[utf8]{inputenc}
\usepackage[T1]{fontenc}
\usepackage{mathptmx}
\usepackage{textgreek}
\usepackage{upgreek}
\usepackage{float}
\usepackage{hyperref}

\usepackage{color}

\usepackage{natbib}

\makeatletter
\def\maketag@@@#1{\hbox{\m@th\normalfont\normalsize#1}}
\makeatother

\newcommand{\beginsupplement}{%
        \setcounter{table}{0}
        \renewcommand{\thetable}{S\arabic{table}}%
        \setcounter{figure}{0}
        \renewcommand{\thefigure}{S\arabic{figure}}%
				\renewcommand{\theequation}{S.\arabic{equation}}
     }

\begin{document}

	\preprint{}

\title[MCA for NbSe2]{Supercurrent diode effect and magnetochiral anisotropy in few-layer NbSe$_2$} 
% Force line breaks with \\

\author{Lorenz Bauriedl}
\author{Christian B\"auml}
\author{Lorenz Fuchs}
\author{Christian Baumgartner}
\author{Nicolas Paulik}
\author{Jonas M.~Bauer}
\author{Kai-Qiang Lin}
\author{John M.~Lupton}
\affiliation{Institut f\"ur Experimentelle und Angewandte Physik, University of Regensburg, Regensburg, Germany}
\author{Takashi Taniguchi}
\author{Kenji Watanabe}
\affiliation{International Center for Materials Nanoarchitectonics, National Institute for Materials Science, Tsukuba, Japan}
\author{Christoph Strunk}
\author{Nicola Paradiso}\email{nicola.paradiso@physik.uni-regensburg.de}
\affiliation{Institut f\"ur Experimentelle und Angewandte Physik, University of Regensburg, Regensburg, Germany}


\begin{abstract}	
Nonreciprocal transport refers to charge transfer processes  that are sensitive to the bias polarity. Until recently, nonreciprocal transport was studied only in dissipative systems, where the nonreciprocal quantity is the resistance. Recent experiments have, however, demonstrated nonreciprocal supercurrent leading to the observation of a supercurrent diode effect in Rashba superconductors, opening the vision of dissipationless electronics. 
Here we report on a supercurrent diode effect in NbSe$_2$ constrictions obtained by patterning NbSe$_2$ flakes with both even and odd layer number. The observed rectification is driven by valley-Zeeman spin-orbit interaction. We demonstrate a rectification efficiency as large as 60\%, considerably larger than the efficiency of devices based on Rashba superconductors. In agreement with recent theory for superconducting transition metal dichalcogenides, we show that the effect  is driven by an out-of-plane magnetic field component. Remarkably, we find that the effect becomes field-asymmetric  in the presence of an additional in-plane field  component transverse to the current direction. Supercurrent diodes offer a further degree of freedom in designing superconducting quantum electronics with the high degree of integrability offered by van der Waals materials. 
\end{abstract}

\maketitle

The archetypal example of a nonreciprocal electronic device is the diode. The term nonreciprocity in this context is used to imply a large difference in resistance between opposite bias polarities. For a conventional semiconductor diode, nonreciprocity follows from the inequivalence of the two crystals forming the $pn$ junction, which have different types of doping. In \textit{homogeneous} devices, polarity-dependent resistance is observed when both inversion and time-reversal symmetry are  broken simultaneously. As shown by Rikken \textit{et al.}~\cite{RikkenPRL2001,RikkenPRL2005}, in noncentrosymmetric conductors nonreciprocal resistance can be phenomenologically described by 
%in the most general case the Onsager relations predict the following magnetic field correction to the resistance
\begin{equation}
    R=R_0[1+\alpha B^2 + \gamma BI],
    \label{eq:rikken}
\end{equation}
where the coefficient $\alpha$ refers to the usual magnetoresistance and  $\gamma$ is the magnetochiral anisotropy (MCA) coefficient. In normal conductors, $\gamma$  is typically very small. Its strength is, in fact, determined by the ratio between spin-orbit perturbation and the Fermi energy. Nonreciprocal transport therefore becomes discernible in semiconductors with low Fermi level and large spin-orbit interaction (SOI)~\cite{Ideue2017}. The MCA effect can be greatly amplified in noncentrosymmetric superconductors~\cite{Wakatsuki2017,Yasuda2019,Itahashi2020,Ideue2020,Zhang2020,Itahashi2020}. Here, the energy scale governing the fluctuation regime close to the critical temperature is not the Fermi energy, but the superconducting gap. Another way to boost MCA for the resistance is to engineer nonreciprocal vortex motion in a superconductor by asymmetric patterning of artificial pinning centers~\cite{Lyu2021}.

In the past years, superconductivity-enhanced MCA for the resistance has been studied extensively. There is a satisfactory understanding of the mechanisms producing nonreciprocal resistance and theory predictions successfully describe the experiments~\cite{Hoshino2018,Tokura2018}. On this basis, some first applications, such as superconducting antenna rectifiers~\cite{Zhang2020} or spin filtering diodes~\cite{strambini2021arx}, have already been proposed.

\begin{figure*}[tb]
\includegraphics[width=\textwidth]{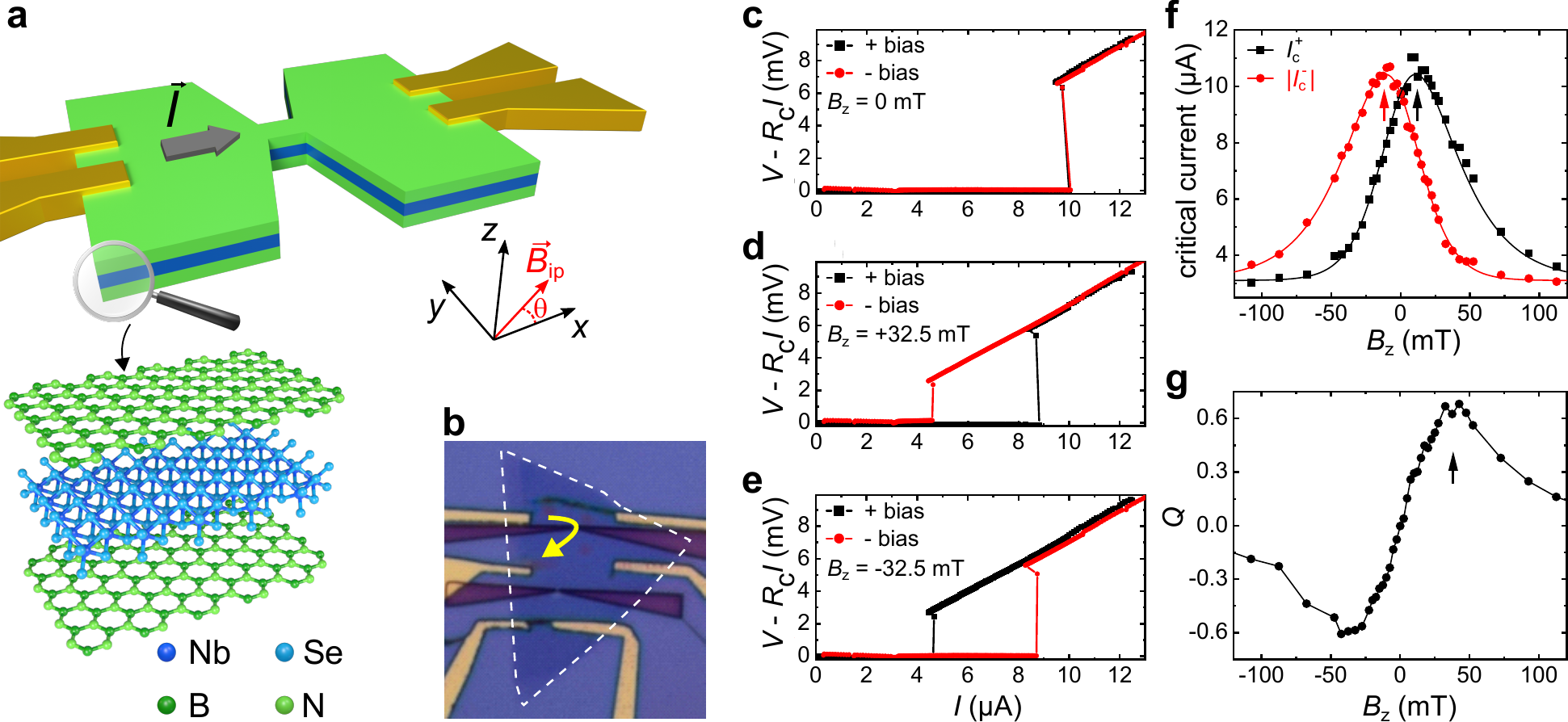}
\caption{\textbf{Supercurrent diode effect in a van der Waals superconductor.} \textbf{a}, Scheme of the typical device. The central constriction is 250~nm-wide and 250~nm-long. The $x$-direction is chosen to be that of the supercurrent, i.e., the constriction  axis. The $z$-direction is perpendicular to the crystal plane. The device is fabricated starting from a stack of hBN ($\approx 10$~nm, tens of layers), NbSe$_2$ (2, 3 or 5 layers) and again hBN (tens of layers). \textbf{b}, Optical micrograph of sample B. The dashed white contour highlights the NbSe$_2$ crystal. Electrodes are fabricated by edge contact techniques~\cite{baeuml2021CNT,HuangCNTedge,Wang614}, while the constrictions are made by reactive ion etching. The yellow arrow indicates the supercurrent pathway. 
\textbf{c},~Current-voltage characteristics (IVs) for opposite bias polarities (i.e., opposite current directions) measured on sample G in a 3-terminal configuration for zero out-of plane field  $B_z$. The sweep direction is always from zero to finite bias. A contact resistance $R_c=1$~k\textOmega ~has been subtracted. \textbf{d},~Similar measurements, but for $B_z=32.5$~mT. Notice the difference between the two critical currents. \textbf{e},~Same as in panel \textbf{d}, but with opposite field orientation. Notice that the role of the two bias polarities now is swapped. \textbf{f}, Absolute critical current for positive (black) and negative (red) bias as a function of $B_z$. Each value is the average of 10 consecutive measurements. The critical current is maximal for a nonzero $B_z$, namely for $|B_z|=B_{\text{max},I_c}\approx 10$~mT (black and red arrow). The red and black solid line are guides to the eye, mutually symmetric upon reflection around $B_z=0$. \textbf{g}, Supercurrent rectification efficiency $Q\equiv 2(I_c^+-|I_c^-|)/(I_c^++|I_c^-|)$, plotted versus $B_z$. $Q$ is maximal for $B_z=B_{\text{max},Q}\approx 35$~mT (arrow). Measurements in \textbf{c--g} were performed at 1.3~K. As discussed in the Supplementary Information, an offset of $-2.5$~mT and 0.17~\textmu A has been removed from $B_z$ and $I_c$, respectively.
}
\label{fig:firstfig}
\end{figure*}

Until recently, studies on nonreciprocal transport, even when exploiting superconductivity to enhance MCA, have focused exclusively on resistance. The seminal demonstration of a dissipationless nonreciprocal \textit{supercurrent} \cite{HuPRL2007} was only recently  provided by experiments on synthetic Rashba superconductors based on Nb/V/Ta multilayers~\cite{Ando2020} and on Josephson junctions~\cite{baumgartner2021diode} with strong Rashba SOI. Such a nonreciprocal supercurrent gives rise to the so-called superconducting diode effect, where the  supercurrent can flow only in one direction that can be switched by a magnetic field~\footnote{Since the terminology has not settled yet in the literature, it is possible to find the term \textit{superconducting diode effect} referred to resistance MCA, when this latter is enhanced by superconducting fluctuation near the critical temperature. Here, we refer exclusively to the (dissipationless) supercurrent diode effect.}.

The MCA for the electrical resistance and that for the superflow are two clearly distinct effects.
The latter is characterized by the kinetic inductance, which is uneven in the current, while the resistance is zero. In the experiments reported so far, samples that showed nonreciprocal supercurrent far below $T_c$ also showed nonreciprocal resistance in the fluctuation regime near $T_c$~\cite{Ando2020,baumgartner2021diode}.  This correlation between the two phenomena arises due to the fact that both effects require the same symmetry breaking mechanisms (i.e., time and inversion symmetry). While  there is a satisfactory understanding of MCA for the resistance, the theoretical study of magnetochiral effects in supercurrents is just in its infancy~\cite{yuan2021pnas,daido2021prl,he2021njp,Ilic2022,halterman2021supercurrent,Fu2022,He2022,scammell2021theory,zhang2021general,zinkl2021symmetry}. Nonreciprocal supercurrent is better understood in Josephson junctions, where the diode effect can be engineered by Andreev bound states in the normal weak-link. %In this case, the diode effect can be explained in terms of even (cosine) components added to the current-phase relation by applying an in-plane  magnetic field. The mechanism is less clear for bulk samples, where a consolidated theory of the supercurrent diode effect does not exist yet, although several models have been proposed recently~\cite{yuan2021arx,daido2021prl,he2021arx}.
The observation of non-reciprocal supercurrent in Rashba superconductors raises the question about its existence in materials with other types of SOI. Promising candidates are transition metal dichalcogenides (TMDs) that feature valley-Zeeman SOI, where magnetochiral resistance was already   measured, while non-reciprocal supercurrent was predicted~\cite{he2021arx} but not yet observed.  

Here, we report on the observation of a pronounced supercurrent diode effect in constrictions of few-layer NbSe$_2$-crystals. Owing to dominant valley-Zeeman SOI, the supercurrent diode behavior is driven by the out-of-plane component $B_z$ of the magnetic field. Unexpectedly, also the in-plane component affects the non-reciprocal supercurrent: 
we find that it breaks the anti-symmetry of the critical current difference with respect to $B_z$, boosting the rectification for one $B_z$ direction, and suppressing it for the opposite one.
%Near $T_c$, the supercurrent rectification efficiency follows the $\sqrt{1-T/T_c}$-behavior predicted in~\cite{he2021arx}; \textcolor{red}{at lower temperatures it displays a non-monotonic behavior with a maximum near $T_c/2$.}

Our samples are fabricated using standard exfoliation methods for TMDs~\cite{Wang614,Castellanos_Gomez_2014,Paradiso_2DMat_2019,Holler_2019}. A scheme of the typical device is depicted in Fig.~\ref{fig:firstfig}\textbf{a}. The exfoliated NbSe$_2$ crystals considered here are between 2 and 5 layers thick~\cite{Paradiso_2DMat_2019}. The flake thickness can be estimated with reasonable accuracy from the optical contrast of the crystal when stamped on standard SiO$_2$/Si substrates. The parity of the layer number $N$ and the lattice orientation is determined, \textit{a posteriori},  by second harmonic generation (SHG) measurements for all samples. NbSe$_2$ crystals are fully encapsulated in hBN~\cite{Holler_2019} and edge contacts are fabricated by electron beam lithography~\cite{baeuml2021CNT}. A 250~nm-long channel of  250~nm width is patterned by electron beam lithography and reactive ion etching. The etched parts appear as dark  purple triangles in Fig.~\ref{fig:firstfig}\textbf{b}. The purpose of the narrow channel is to have a well-defined direction of the current density in the constriction, direction indicated by $\vec{I}$ in Fig.~\ref{fig:firstfig}\textbf{a}. In what follows, the $x$-direction is assumed to be that of the constriction axis (and thus that of the supercurrent) while the $z$-direction is perpendicular to the sample plane.

The main measurements of this work (sample F, see below) were performed in a dilution refrigerator equipped with a central superconducting coil controlling the in-plane component of the magnetic field. Additional coils provide a field perpendicular to the sample plane. A piezo-rotator allows for rotation of the sample around an axis normal to the sample plane. 
%make possible to rotate the sample, which is equivalent to rotate the direction of the in-plane field with respect to the NbSe$_2$ constriction axis.
Additional measurements (samples B--E and G, see Supplementary Information)  were performed in a $^4$He cryostat with a base temperature of 1.3~K, equipped with a single superconducting coil. 
%The field in this case is oriented in-plane, but with a non-negligible  out-of-plane component due to a small sample misalignment of a few degrees.
Figures~\ref{fig:firstfig}\textbf{c-e} introduce the supercurrent diode effect as measured in a NbSe$_2$ constriction patterned on sample G.
%, which is depicted in Fig.~\ref{fig:firstfig}\textbf{b}
The graphs show three pairs of current-voltage characteristics (IVs) for applied out-of-plane magnetic fields of $B_z=0$~mT (\textbf{c}), $B_z=32.5$~mT (\textbf{d}) and $B_z=-32.5$~mT (\textbf{e}). Importantly, all the IVs reported here always refer to the  zero-to-finite (either positive or negative) bias sweep direction, in order to rule out heating effects. In the three graphs, the black (red) symbols refer to current density in the narrow channel oriented towards the positive (negative) $\hat{x}$ direction. A strong supercurrent diode effect~\cite{Ando2020,baumgartner2021diode} is evident from the comparison of the IVs: there is a marked difference $\Delta I_c\equiv I_c^+-|I_c^-|$ in the critical current for the two supercurrent orientations, the sign of which change when the magnetic field $B_z$ is inverted.
The complete field dependence of both $I_c^+$ (black) and $|I_c^-|$ (red) is plotted in Fig.~\ref{fig:firstfig}\textbf{f}.
The current range between $I_c^+$ and $|I_c^-|$ corresponds to the supercurrent diode regime, where current flows without dissipation only in one direction, which can be selected by changing the sign of the magnetic field~\cite{Ando2020}. As a figure of merit of the supercurrent  diode effect one can take the supercurrent  rectification efficiency $Q\equiv 2\Delta I_c/(I_c^++|I_c^-|)$~\cite{he2021njp}, i.e.~the difference between $I_c^+$  and $I_c^-$  normalized by their average, which is plotted in Fig.~\ref{fig:firstfig}\textbf{g} versus $B_z$. For moderate fields, $Q$ increases almost linearly. The maximum magnitude of  $Q$ is above 60\%, much larger than that ($\approx 5$\%) observed in Ref.~\cite{Ando2020}.  Beyond a certain breakdown field $B_{\text{max},Q}\approx 35$~mT the diode effect is gradually suppressed. This behavior is reminiscent of that observed in  Rashba superconductors, see e.g.~Fig.~2 in Ref.~\cite{Ando2020} and Fig.~3 in Ref.~\cite{baumgartner2021diode}.   Theory models for Rashba superconductors~\cite{yuan2021pnas,Ilic2022} predict a similar suppression, but the threshold field is expected to be of the order of the paramagnetic limit, i.e.~much larger than the value observed in our and in literature experiments~\cite{Ando2020,baumgartner2021diode}. Finally, we stress that the observations in Fig.~\ref{fig:firstfig}\textbf{f,g} do not depend on the field sweep direction, as discussed in the Supplementary Information.

It is important to note that in our case the diode effect can be so strong that, for one of the two bias polarities, the critical current \textit{increases} by increasing the magnetic field, reaching a maximum at nonzero field. This can be seen, e.g., in Fig.~\ref{fig:firstfig}\textbf{f} for $|B_z|<B_{\text{max},I_c}\approx 10$~mT.
The remarkable increase of the critical current (the origin of which certainly deserves further study) is important since it eliminates the possibility that the nonreciprocal supercurrent originates from nonreciprocal Joule heating due to, e.g.,  vortex motion through asymmetric barriers. In the latter case, the diode effect could result from a different rate of vortices traversing the constriction in the two transverse directions. Such different vortex creep rate would imply a different Joule dissipation, a different electron temperature and finally a different critical current for the two bias polarities. In contrast to our observations,  the maximum critical current would then be always observed in absence of vortices, i.e., at $B_z=0$. The irrelevance of vortices in our experiment is further discussed in the Supplementary Information.

%This fact excludes the possibility  that the nonreciprocal supercurrent (the new effect to be demonstrated here) is just an artifact of nonreciprocal \textit{resistance}, which in turn is well-studied in superconducting  TMDs~\cite{Wakatsuki2017,Yasuda2019,Itahashi2020,Ideue2020,Zhang2020,Itahashi2020}. In principle, if the resistance is finite and  polarity-dependent, at finite current the dissipated power would also be asymmetric, producing a different  suppression of the critical current  for the two directions. This difference could potentially  mimic a supercurrent diode effect. The enhancement of the  critical current observed in the present experiment unambiguously rules out this hypothesis. 

%Interestingly, the relative increase of the critical current is sample dependent and not necessary for the observation of the diode effect: sample D and F both show a pronounced diode effect, yet the field-induced increase of the critical current is  much smaller than that observed in Fig.~\ref{fig:firstfig}\textbf{g} (sample D) or nearly absent (sample F)~\footnote{See Supplementary Materials for further details.}.

\begin{figure*}[tb]
\includegraphics[width=2\columnwidth]{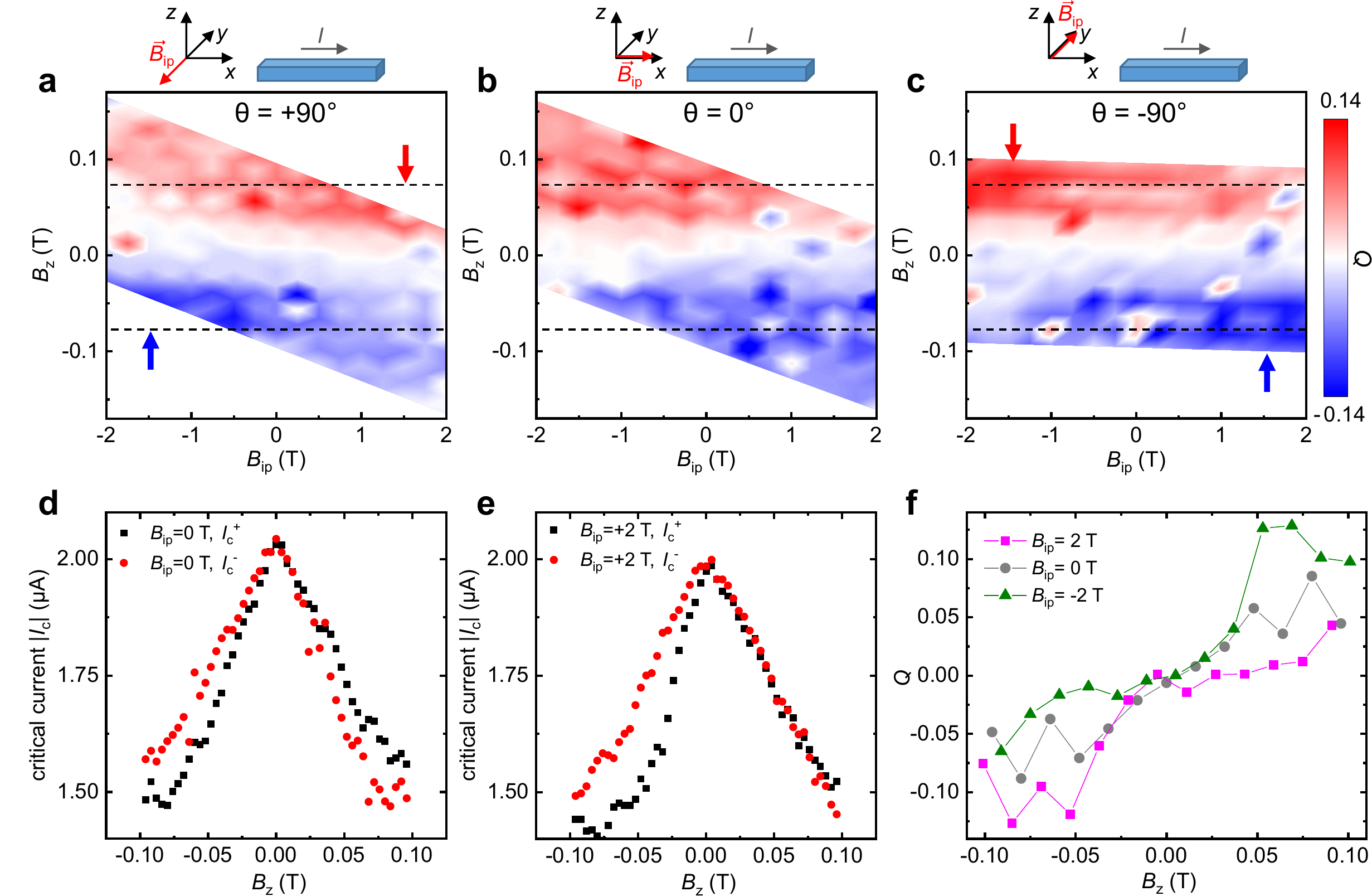}
\caption{\textbf{Supercurrent diode effect versus in- and out-of-plane field.} \textbf{a}, The color plot shows $Q \equiv 2( I_c^+ - |I_c^-|)/(I_c^+ + |I_c^-|)$ as a function of out-of-plane ($B_z$) and in-plane ($B_{ip}$) field, measured in sample F for $\theta = 90^{\circ}$ (i.e., for $\vec{B}_{ip}\perp \vec{I}$). $B_z$ here includes both the field produced by the orthogonal coils and the finite $z$-component of $\vec{B}_{ip}$ arising due to misalignment. Red and blue arrows indicate the areas where the diode effect is enhanced. \textbf{b}, Similar measurements, but for $\theta=0^{\circ}$.   \textbf{c}, As in \textbf{b}, but for $\theta=-90^{\circ}$. Notice that this graph can be mapped onto that in \textbf{a}, provided that $B_{ip} \rightarrow -B_{ip}$. 
The color contrast is the same in \textbf{a}, \textbf{b} and \textbf{c}. \textbf{d}, Absolute value of $I_c^+$ (black) and $I_c^-$ (red) plotted versus $B_z$, for $B_{ip}=0$ and for the sample orientation  $\theta=-90^{\circ}$. \textbf{e},~As in \textbf{d}, but for $B_{ip}=2$~T. Note that data in  \textbf{d},\textbf{e} were measured in a different session (with higher resolution in $B_z$) compared to data in \textbf{c}. \textbf{f},~Supercurrent rectification efficiency $Q$ as a function of $B_z$ at $\theta=-90^{\circ}$ for $B_{ip}=-2$~T  (green), 0~T (gray), and 2~T (magenta). Here, we used the same data as in panel \textbf{c}, for $B_{ip}$ values indicated in the legend.
%To reduce the data scatter, data are averaged over adjacent values of $B_{ip}$: data for $B_{ip}=\pm 2$~T are the average of the curves for $\pm 1.75$ and $\pm 2$~T, while data for $B_{ip}=0$~T are obtained by averaging the   curves for -0.25, 0 and 0.25~T. 
For $B_{ip}=0$ we have substituted three outliers (for $B_z = -48$, $-64$ and $-80$~mT) with the corresponding values for the adjacent in-plane field $B_{ip}=-0.25$~T. For $B_{ip}=-2$~T we have substituted one outlier  (for $B_z = -43$~mT) with the corresponding value for the adjacent in-plane field $B_{ip}=-1.75$~T, see Supplementary Information. Outliers are also visible in panels \textbf{a}-\textbf{c}. 
All measurements reported in this figure were performed at 500~mK.
}
\label{fig:mainresults}
\end{figure*}

%We find that the relative asymmetry for this first samples is higher than that found in bulk superlattices REFE TO ANDO. As shown below, the values is even larger for another samples, discussed below. 

We observed supercurrent rectification in other devices with the same nominal geometry. The supercurrent diode effect was clearly visible in all the samples where supercurrent could be measured, i.e., sample B, D, E, F, and G. 
Their layer number $N$ was determined by combining information from white-light  optical and SHG microscopy~\cite{Xi2015Ising,MalardSHG13,LiSHG13,Lin2019}, as discussed in the Supplementary Information. We found a layer number $N=3$ for sample B, D and F, $N=5$ for sample E and $N=2$ for sample G.
As for the Ising superconductivity, the supercurrent diode effect is not theoretically expected to occur in NbSe$_2$ when $N$ is even, owing to the restored inversion symmetry. However, both Ising superconductivity (see Fig.~4 in Ref.~\cite{Xi2015Ising}) and  the supercurrent diode effect (this work) are experimentally observed for both even and odd $N$. This fact is likely because of the relatively weak electronic coupling  between the layers, which effectively renders NbSe$_2$ a collection of monolayers~\cite{Xi2015Ising}.

%On the one hand, a  large number of samples of either layer parity would be necessary to establish a correlation between layer number parity and visibility of the supercurrent diode effect. On the other, our observation at least suggest that inversion symmetry needs to be broken (as in odd-layer 2H-polytype of NbSe$_2$)  in order to observe magnetochiral effects. This is expected since broken inversion symmetry enables SOI, which  (together with field induced breaking of time-reversal symmetry) ultimately produces the supercurrent  diode effect.

%The measurements on sample B  (and on samples D and E, which show similar results) performed in the 1~K $^4$He cryostat cannot disentangle the effects of the different components  of the magnetic field, i.e., in- and out-of-plane. 

The experimental results shown in Fig.~\ref{fig:firstfig} clearly demonstrate that the out-of-plane field is required for the supercurrent diode effect in materials with valley-Zeeman spin-orbit. However, it is interesting to study whether the in-plane field also affects the rectification and, if it does, what the role of the components perpendicular and parallel to the current may be.
For this reason, we measured one device, sample F, in a setup equipped with a piezo-rotator.
This setup allows us to rotate the sample with respect to the main magnetic field $\vec{B}_{ip}$, and therefore to control the angle $\theta$ between $\vec{B}_{ip}$ and supercurrent $\vec{I}$, as indicated in Fig.~\ref{fig:firstfig}\textbf{a}. Additional coils, perpendicular to the main one, provide an independent control of the out-of-plane field $B_z$. 
Figures~\ref{fig:mainresults}\textbf{a-c} show the supercurrent rectification efficiency $Q$ in sample F, plotted as a function of both the in-plane (${B}_{ip}$) and the out-of-plane ($B_z$) magnetic field, for $\theta=90^{\circ}$ (panel \textbf{a}, $\vec{B}_{ip}$ transverse to the supercurrent), for $\theta=0^{\circ}$ (panel \textbf{b}, $\vec{B}_{ip}$ parallel to the supercurrent), and for  $\theta=-90^{\circ}$ (panel \textbf{c}, $\vec{B}_{ip}$ antiparallel to that in panel \textbf{a}). Owing to sample misalignment, a large in-plane field produces a significant out-of-plane component. This misalignment can be quantified (and accounted for, as in Fig.~\ref{fig:mainresults}\textbf{a-c}) by looking at the maximum of $I_c^+(B_z)$ [or $I_c^-(B_z)$], as discussed in the Supplementary Material. As a result of this offset adjustment, the data range in the $(B_{ip},B_z)$ plane appears as a rhomboid.

The most prominent feature in these measurements is the vertical gradient in the color plot, indicating that the rectification efficiency increases monotonically with $B_z$ and is zero when $B_z=0$, at least within the experimental data scatter. We have therefore now  established experimentally that, as predicted by theory~\cite{he2021arx}, the supercurrent diode effect in TMDs is driven by the out-of-plane field only. In contrast, it is the in-plane field that  drives the diode effect in Rashba superconductors. 
Moreover, unlike what is predicted for superconductors with pure valley-Zeeman SOI~\cite{he2021arx},  we observe that the in-plane field does affect the supercurrent diode effect. More precisely, the in-plane field component \textit{perpendicular to the current} breaks the (anti)symmetry of $Q$ as a function of $B_z$.  If $B_{ip,y}\equiv \vec{B}_{ip}\cdot \hat{y}=0$ then  $Q (B_z)=-Q (-B_z)$, with no dependence on $B_{ip,x}\equiv \vec{B}_{ip}\cdot \hat{x}$, see Fig.~\ref{fig:mainresults}\textbf{b}. Instead, Fig.~\ref{fig:mainresults}\textbf{a},\textbf{c} prove that, for finite $B_{ip,y}$,  $Q (B_z) \neq -Q(-B_z)$, i.e., the diode effect becomes asymmetric in $B_z$. This effect is apparent, e.g., in Fig.~\ref{fig:mainresults}\textbf{a}: for the given field and current orientation, the diode effect is enhanced when $B_z$ and $B_{ip}$ are both positive or negative (first and third quadrant of the graph, deep red and deep blue region, indicated by the arrows). On the other hand, it is suppressed when $B_z$ and $B_{ip}$ have opposite signs (second and fourth quadrant, light red and light blue region). At fixed $B_z$ (dashed lines), inverting the sign of $B_{ip,y}$ corresponds to a transition from enhanced to suppressed diode effect, i.e. from deep blue to light blue for the upper dashed line. Clearly, a $180^{\circ}$ sample rotation is equivalent to a sign change in $B_{ip}$, as it is evident by comparing Fig.~\ref{fig:mainresults}\textbf{a} and Fig.~\ref{fig:mainresults}\textbf{c}.
It is important to remark that data in Fig.~\ref{fig:mainresults}\textbf{a}-\textbf{c} are still (anti)symmetric upon simultaneous inversion of  both $B_z$ and $B_{ip,y}$, i.e., $Q(B_z,B_{ip,y})=-Q(-B_z,-B_{ip,y})$.    

To better visualize the impact of the in-plane field, it is instructive to look separately at $I_c^+$ and $I_c^-$ as a function of $B_z$, both with and without in-plane field. The latter case is shown in Fig.~\ref{fig:mainresults}\textbf{d}. Both $I_c^+(B_z)$ and $I_c^-(B_z)$ appear as asymmetric, $\Lambda$-shaped functions. Their difference in slope for positive and negative $B_z$ produces a supercurrent diode effect  increasing monotonically  with $B_z$.
We note the similarity of Fig.~\ref{fig:mainresults}\textbf{d} and Fig.~\ref{fig:firstfig}\textbf{f} with the corresponding measurements on Rashba Josephson junctions, see Fig.~3\textbf{e} in Ref.~\cite{baumgartner2021diode}.
As for Fig.~\ref{fig:firstfig}\textbf{g}, the whole graph in Fig.~\ref{fig:mainresults}\textbf{d} is symmetric if both the sign of $B_z$ and the direction of the supercurrent are changed simultaneously, i.e.~for $B_z \leftrightarrow -B_z$ and $I_c^+ \leftrightarrow I_c^-$. 
This symmetry is broken when an in-plane field is applied along the direction perpendicular to the supercurrent. In the case of $B_{ip,y}>0$. shown in Fig.~\ref{fig:mainresults}\textbf{e}, the difference between the slope of $I_c^+(B_z)$ and $I_c^-(B_z)$ is reduced for $B_z>0$ and enhanced for $B_z<0$, and vice versa for $B_{ip,y}<0$. 

Figure~\ref{fig:mainresults}\textbf{f} summarizes our main findings: the supercurrent rectification efficiency $Q$ is plotted as a function of $B_z$ for $\theta=-90^{\circ}$ and for $B_{ip}=2$~T, 0~T and -2~T. The graph shows that the perturbation due to the in-plane field is evident only for sufficiently high out-of-plane field, i.e., $|B_z|>30$~mT. Above that threshold, for $B_z>0$, the slope of $Q$ versus $B_z$ increases strongly for $B_{ip}=-2$~T and decreases for  $B_{ip}=2$~T. The opposite is true for $B_z<0$. We also note that at sufficiently large $|B_z|$, above approximately  50~mT, the diode effect starts to be suppressed, so that at around $|B_z|\gtrsim 100$~mT the different curves in Fig.~\ref{fig:mainresults}\textbf{f} merge again.
 
The key result of our observations is that the supercurrent diode effect in NbSe$_2$ is controlled by the out-of-plane field, as predicted by theory for superconducting TMDs~\cite{he2021arx}. For $B_z=0$ \textit{there is no diode effect}, independent of $\vec{B}_{ip}$. On the other hand, if $\vec{B}_{ip}$ has a component $B_{ip,y}$ perpendicular to the current, then   the supercurrent diode effect becomes \textit{asymmetric} in $B_z$, i.e., it is enhanced for one out-of-plane field polarity and suppressed for the other. The role of the two polarities is swapped if both the sign of the supercurrent and the sign of $B_{ip,y}$ are inverted. To the best of our knowledge, there are no predictions to date regarding a possible role of the in-plane field on the supercurrent diode effect in TMDs. On the contrary,  for Rashba superconductors it is precisely the in-plane field that controls MCA and the diode effect~\cite{Ando2020,baumgartner2021diode}. It is clear that further studies are needed to elucidate the role of the in-plane field in superconducting TMDs, which might be related to a possible Rashba-like component of the SOI~\cite{Lu1353,He2018CP}. In NbSe$_2$ such Rashba components can arise, e.g., due to ripples in the  crystal~\cite{Huertas2006} formed when stamping NbSe$_2$ on hBN~\cite{Holler_2019}, or due to the substrate~\cite{Hamill2021}. 
The existence of a weak Rashba SOI component in NbSe$_2$ has, for example, been invoked to explain the apparent two-fold anisotropy of the magnetoresistance as a function of in-plane magnetic field~\cite{Hamill2021}. The model proposed in Ref.~\cite{Hamill2021} relies on the presence of $p$-wave components in the pairing function, which might as well play a role in the anisotropy of the diode  effect we observe here.

\begin{figure}[tb]
\includegraphics[width=1\columnwidth]{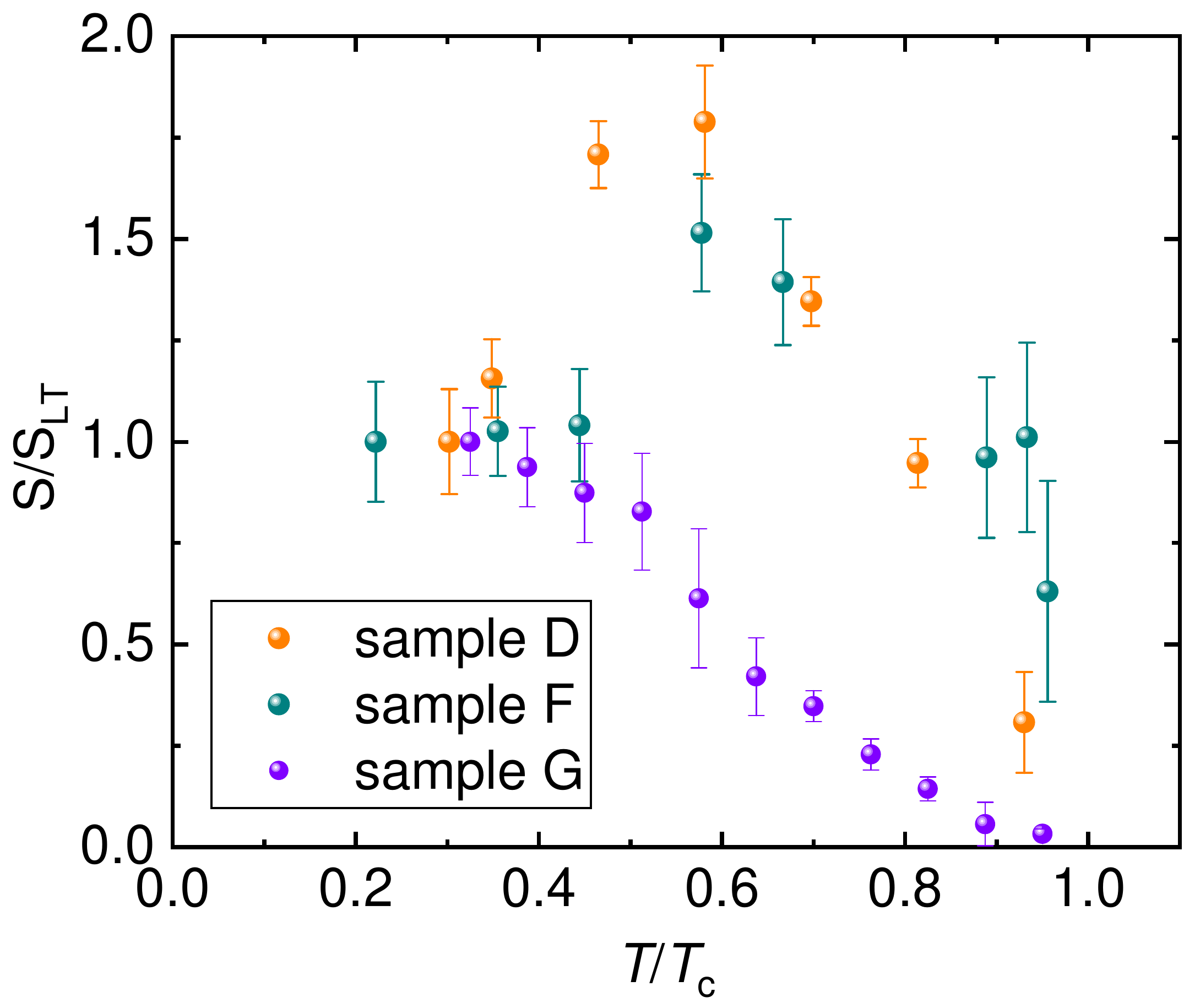}
\caption{\textbf{Temperature dependence of the supercurrent diode effect.} The graph shows the temperature dependence of  $S\equiv Q/B_z$ for sample D, F and G. $S$ is  evaluated from $Q$ near $B_z=0$, i.e.~in the linear regime of the supercurrent rectification efficiency, where $Q\propto B_z$. The $S$ values are normalized to the lowest temperature value $S_{\text{LT}}$ (e.g.~17.2~T$^{-1}$ for sample G),
while the temperatures are normalized to the critical temperature $T_c$ of the corresponding sample. Orange, blue  and purple symbols refer to sample D, F and G, where the critical temperature of the constriction region is 4.3~K, 2.25~K and 4.0~K, respectively.} 
\label{fig:tempdep}
\end{figure}

Next, we turn to the temperature dependence of $Q$. 
In the literature, theory predictions regarding the temperature dependence of the rectification efficiency are quite diverse. While some models predict a square-root-like dependence near the critical temperature~\cite{he2021njp}, others suggest a more complex functionality~\cite{Ilic2022}.
%In Ref.~\cite{he2021arx} a $Q \propto \sqrt{1-T/T_c}$ dependence was predicted for TMD superconductors based on Ginzburg-Landau theory, which is thus strictly valid only near $T_c$. 
Figure~\ref{fig:tempdep} shows the temperature dependence of the diode effect in samples D,  F and G. In the graph we plot $S\equiv Q/B_z$, obtained from $Q$ in the linear regime of low field, where $Q\propto B_z$.  We choose to display $S$ rather than $Q$ because it represents a linear interpolation of $Q(B_z)$ and thus  it averages over several data points, therefore showing a reduced scatter compared to $Q$ for a given $B_z$.

First of all, we observe that the supercurrent diode effect saturates at low temperature. This saturation is compatible with all theory models and similar to the results on Josephson junctions in Rashba superconductors~\cite{baumgartner2021diode}. This result is   not obvious, since in the data in Ref.~\cite{Ando2020}  the diode effect is visible only near $T_c$ and it is suppressed for both higher and lower temperatures. We note also that the temperature dependence of the diode effect in sample D and F is nonmonotonic, with a maximum reached near $T/T_c=0.5$, where, instead, a kink is observed in sample G. 
A nonmonotonic temperature dependence  of the rectification at finite field was also predicted by recent theory~\cite{Ilic2022}. It is possible that the different behavior displayed by samples D, F and G is due to the fact that, in order to obtain a significant rectification, we need to apply a sufficiently strong field. Notice that theory in Ref.~\cite{Ilic2022} predicts that the temperature dependence of the rectification is critically affected by a finite magnetic field.
%We stress that this measurement becomes difficult near $T_c$, since both $I_c^+$ and $I_c^-$ tend to zero and it is thus difficult to measure a relatively small difference between them. While we can experimentally confirm the  $\sqrt{1-T/T_c}$ dependence (dashed gray line in Fig.~\ref{fig:tempdep}) of the rectification efficiency $Q$ predicted by theory, its quantitative verification   is challenging near $T_c$.

A comment is in order about features which appear to be sample dependent. While the supercurrent diode effect was observed in every sample we measured (see Supplementary Information for further details), the maximum rectification efficiency is highly scattered among the samples, ranging from 6 to 60\%. Also the supercurrent  increase with $B_z$ is sample dependent, both in terms of relative increase  $I_c(B_{\text{max},I_c})/I_c(0)$ and maximum supercurrent field $B_{\text{max},I_c}$. The supercurrent increase seems to be independent from the efficiency of the supercurrent rectification:  the former is negligible  in sample F, while in sample G no particular feature is observed in $Q(B_z)$ at $B_z=B_{\text{max},I_c}$.
The nominal differences between the samples are the layer number and the orientation of $\vec{I}$ with respect to the lattice. Both features can be determined by SHG microscopy performed after the transport measurements, as described in the Supplementary Information. We found that neither the layer number nor the supercurrent-to-lattice orientation are correlated with the magnitude of the rectification efficiency. On the other hand, 
lithography of narrow constrictions by reactive ion etching  produces random disorder, in particular at the edges. This randomness, together with the still limited number of studied samples, does not allow us to conclusively assess the role of the lattice orientation.
%A comment is in order about the difference in field dependence of the critical current [either $I_c^+(B)$ or $I_c^-(B)$] observed in the different samples, see,  e.g., Fig.~\ref{fig:firstfig}\textbf{f},  Fig.~\ref{fig:mainresults}\textbf{d} and Supplementary Information. While all samples measured show a supercurrent diode effect, all but sample F reach their maximum critical current at a finite field, i.e., the $\Lambda$-shaped $I_c^+(B)$  [$I_c^-(B)$] curve reaches a maximum for a positive [negative] field. This maximum critical current  is several percentage points higher than the zero field value, see Fig.~\ref{fig:firstfig}\textbf{f}. The two $I_c^+(B)$ and $I_c^-(B)$ curves  then appear  displaced horizontally. 
%The nominal differences between the samples are the layer number and the orientation of $\vec{I}$ with respect to the lattice (e.g., the armchair direction). Both features can be determined by SHG microscopy performed after the transport measurements, as described in the Supplementary Information. We found that (i) all the measured samples have an odd number of layers and (ii) in sample F, the supercurrent  $\vec{I}$ is directed almost perfectly along the zig-zag direction. 
Further investigation is needed in order to elucidate how the supercurrent diode effect is influenced by the direction of current flow with respect to the crystal.

%Whether this alignment is responsible of fact that sample F has $B_{max}\approx 0$ is something that requires further experimental and theoretical investigation.

%Finally, it is worth mentioning a possible magnetic-field free diode effect in NbSe$_2$-based Josephson junctions~\cite{wu2021}. Whatever its underlying mechanisms ---several possible hypotheses are presented in Refs.~\cite{wu2021,Misaki2021}---, such an observation is, by definition, not a MCA effect. Its physics does not appear to be related to that of the field-controlled supercurrent diode effect studied here and in Refs.~\cite{Ando2020,baumgartner2021diode,he2021njp,yuan2021pnas,daido2021prl}.

In conclusion, we have demonstrated a supercurrent diode effect in  few-layer NbSe$_2$. We show that the effect is controlled by the out-of-plane magnetic field, in contrast to what has been observed in Rashba superconductors, where the effect is driven by the in-plane field component directed perpendicular to the supercurrent. This field component  nevertheless plays a role in NbSe$_2$ devices as well, since it suppresses the diode effect for one out-of-plane field polarity and enhances it for the opposite one. 
Finally, the temperature dependence of the effect shows a saturation at low temperature, a maximum or a kink at around $T=T_c/2$ and a suppression near $T_c$.
% Our work marks an important advance in superconducting electronics, as it demonstrates how non-dissipative diodes can be implemented in devices based on van der Waals materials.
TMD-based diodes can be crucial components in fully superconducting electronics. Their dissipation-free directional transport makes them suited for logic elements, ultrasensitive detectors, or signal demodulators, which can operate at low temperature with no energy loss. Being just a few atoms thick, their performance could conceivably be controlled electrically by gates, and modified by integrated them into complex van der Waals stacks in combination with other 2D materials.

During the review process we became aware of a related experimental work on supercurrent rectification in NbSe$_2$/CrPS$_4$ heterostructures~\cite{shin2021}, in NbSe$_2$-based Josephson junctions~\cite{wu2021}, and on supercurrent rectification in magic-angle twisted bi-~\cite{diezmerida2021magnetic} and trilayer~\cite{lin2021zerofield} graphene, and Nb-proximitized  NiTe$_2$~\cite{palparkin2021}.

%TMD-based diodes can be crucial components in fully superconducting electronics. Their dissipation-free directional transport makes them suited as logic elements, ultrasensitive detectors, or signal demodulators, which can operate at low temperature with no energy loss.

%The dissipation-free directional transport guaranteed by a TMD-based diode opens the way to atomically thin component for fully superconducting electronics. As a diode, it can serve as logic element, usensitive detector, or signal demodulator, but with no energy loss. 

%Being just few atoms-thick, their performances could be controlled electrically by gate, and naturally integrated in complex van der Waals stacks in combination with other 2D materials.

\section*{Methods}
\textbf{Sample preparation.}  NbSe$_2$ crystals were purchased from \textit{HQ Graphene}.
hBN and NbSe$_2$ crystals were exfoliated following two common techniques. One is the technique introduced in Ref.~\cite{Castellanos_Gomez_2014}, where flakes are exfoliated on a polydimethylsiloxane (PDMS) film placed on a glass slide. Suitable crystals are then stamped onto the sample using a micromanipulator placed under a zoom-lens. The other technique is commonly used for the production of fully hBN-encapsulated graphene devices~\cite{Wang614}. In this case,  flakes are sequentially picked up by a thick PDMS film coated with polycarbonate (PC). The flake pick-up takes place at 120$^{\circ}$C, while the final release is triggered by melting the PC at 180$^{\circ}$C and by dissolving it in chloroform. 

Both the fabrication of the contacts and the design of the constriction require an electron beam lithography step followed by reactive ion etching (RIE).  For the RIE, a mixture of 6 sccm O$_2$ and 40 sccm CHF$_3$ is ignited into a plasma with 35~W r.f.~forward power at a pressure of 35~mTorr. The RIE step etches through hBN and NbSe$_2$ with an approximate etching rate of 0.45~nm/s. 
When  electrodes have to be produced, immediately after the RIE step a  10~nm-thick layer of Ti and a 100 nm-thick film of Au are deposited. This procedure reflects the well-known recipe for edge contact fabrication in graphene.
For sample G the RIE process was substituted by an Argon plasma etching process with 2~kV acceleration voltage and 20~mA plasma current at about $3\cdot 10^{-3}$~mbar. The etching rate is in this case approximately 1~nm per minute.

\textbf{Transport measurements.}  
Measurements on sample F were performed in a dilution refrigerator with base temperature 30~mK. For electrical measurements we used DC lines with Cu-powder filters. Since one of the four electrodes stopped working after cool-down, we measured the device in a three-terminal configuration. The IV characteristics were thus obtained by subtracting the voltage $R_cI$, where $R_c=$ 413~\textOmega ~    is the contact resistance.
%The IV characteristics were thus obtained by subtracting a voltage $R_cI$, where $R_c=$ 413~\textOmega ~    is the contact resistance. \textbf{??? to get +-Ic one needs not to substract the offset. In the low temperature limit this offsett would be 413Ohm}. 
In the IV curves, the resistive transition of the constriction at the  critical current was clearly visible as a sharp step, except very close to the critical temperature and field.

The critical currents for data in Fig.~\ref{fig:firstfig} and Fig.~\ref{fig:tempdep} were determined from the IVs as the extrapolation on the current axis (abscissas) of the  steep voltage increase at the resistive transition. In Fig.~\ref{fig:mainresults}, owing to the larger amount of data, we used the alternative criterion $V(I_c)=V_{thres}\equiv 100$~\textmu V, which is better suited for automatic routines. Nevertheless, the  results shown here depend very weakly on the criterion for the critical current. We  verified in all samples that the $I_c^{\pm}$ and  $Q$ data obtained with either criteria were nearly the same (see Supplementary Information for further details).

%for $V=100$~\textmu V. However, the  results shown here depend very weakly on the  criterion for the critical current. As an alternative, we considered the extrapolation to the abscissa, the current axis, of the steep voltage increase at the resistive transition. This criterion has been applied to data for sample B (where the IV curves are noisier due to a resistive voltage electrode), but we have verified in all samples that the $I_c^{\pm}$ and  $Q$ data obtained with either criteria were nearly the same (see Supplementary Information for further details).

Measurements on the other samples (A--E,G) were performed in a   $^4$He cryostat with only room temperature filtering ($\pi$-filters). Except for sample G, the sample holder is positioned in such a way that the magnetic field produced by the superconducting coil is approximately in the plane of the sample surface. The misalignment (typically of the order of a few degrees) produces an out-of-plane field of tens of milliteslas per tesla of applied field. In sample G, instead, the sample is perpendicular to the main field, thus the field is applied exclusively out-of-plane.

\textbf{Second harmonic generation measurements.} Optical measurements of second harmonic generation (SHG)  were always performed \textit{after} transport measurements, in order to minimize the risk of photo-oxidation~\cite{Holler_2019}. The co-polarized SHG intensity is measured as a function of the relative angle between laser polarization and crystal orientation~\cite{Xi2015Ising,MalardSHG13,LiSHG13,Lin2019}.
The monochromatic light source used was a pulsed Ti:sapphire laser (80~fs pulse duration,  80~MHz repetition rate, 1~mW power) at 800~nm. Using a microscope objective (40$\times$, numerical aperture of 0.6), the light was focused onto the NbSe$_2$ samples placed in the vacuum chamber. The reflected SHG signal at 400~nm was collected with the same objective, filtered by a 680~nm short-pass filter, dispersed in a spectrometer (150 grooves/mm grating) and detected by a CCD camera. A linear polarizer was placed in front of the spectrometer to ensure acquisition of the signal polarized  parallel to the laser polarization. A 50:50 non-polarizing beam splitter was used to separate the incident  pathway from the signal detection pathway. In between the beam splitter and the objective, an achromatic half-wave plate was placed to change the relative angle between the crystal orientation and the laser polarization. The half-wave plate was rotated using a stepper motor. A 1200 grooves/mm grating was used as a reference sample. In the experiments, we used an exposure time of 1~s per data point.

\section*{Data and code availability}
The data that support the findings of this study and the code needed to analyze the data are available from the corresponding author upon request.

\section*{Acknowledgments}
\begin{acknowledgments}
\textbf{Acknowledgments.} We thank Denis Kochan, Marco Aprili and Magdalena Marganska for useful discussions. The work was funded by the Deutsche Forschungsgemeinschaft (DFG, German Research Foundation) – Project-ID 
314695032 – SFB 1277 (subprojects B03, B04, B08 and B11), and by the European Union’s Horizon 2020 research and innovation programme
under grant agreements No 862660 QUANTUM E-LEAPS.
\end{acknowledgments}

\section*{Author contributions}
N.~Paradiso and C.~S. conceived and designed the experiments. L.~B., C.~B\"auml, and N.~Paulik fabricated the samples. L.~B., L.~F., C.~Baumgartner and N.Paradiso, performed the transport experiments; L.B., J.~M.~L.,  K.-Q.~L., and J.~M.~B. conceived and performed SHG measurements. K.~W. and T.~T. grew hBN crystals. All authors contributed to the preparation of manuscript.

\section*{Competing interests}
The authors declare  no competing interests.

\section*{Additional Information}
\textbf{Supplementary Materials:} 
Additional measurements on sample G; Measurements on sample B; Summary description of other samples; Second harmonic generation measurements; Role of vortices;  Field misalignment; Resistance and critical current versus temperature; Removal of outliers in Fig.~2f.

\bibliography{biblio}% Produces the bibliography via BibTeX.

\clearpage
\newpage

\onecolumngrid
\begin{center}
\textbf{\large Supplementary Information: Supercurrent diode effect and magnetochiral anisotropy in few-layer NbSe$_2$}
\end{center}

\twocolumngrid
\beginsupplement

\begin{figure*}[tb]
        \centering
        \includegraphics[width=\textwidth]{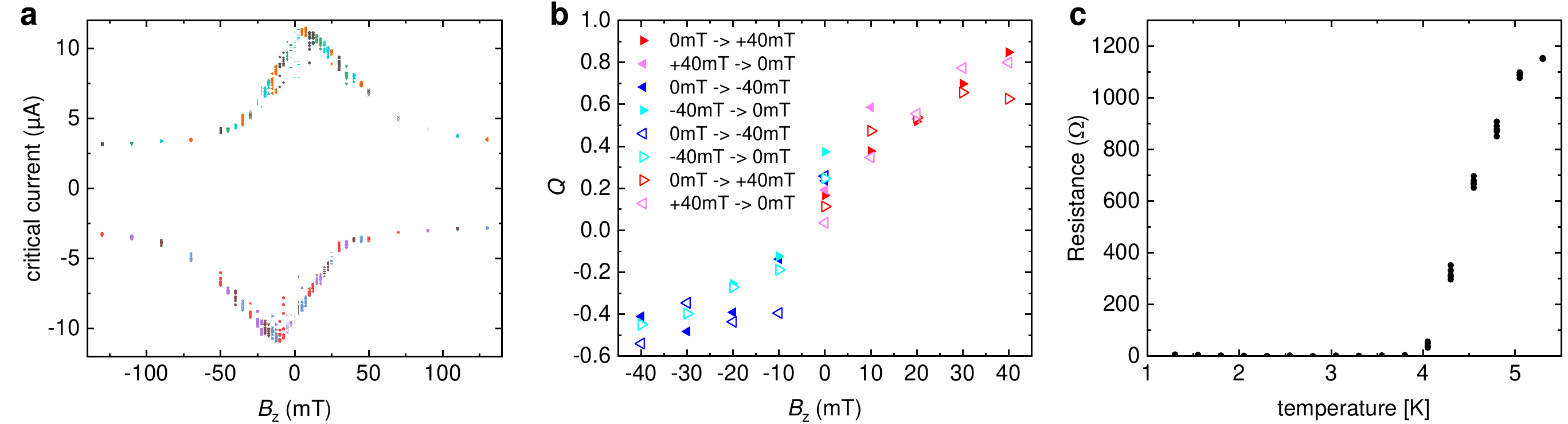}
        \caption{\textbf{Additional measurements on sample G.}  \textbf{a},~Positive and negative critical currents measured as a function of $B_z$ in sample G. For each $B_z$ value the different symbols correspond to different measured IVs. In this case, data are displayed as-measured, without instrumental offset subtraction. The graph highlights the significant spread in the switching currents. \textbf{b},~Supercurrent rectification efficiency $Q$ measured as a function of $B_z$ in sample G. The different symbols and colors refer to different $B_z$-sweep directions. \textbf{c},~Resistance measured in a three-terminal configuration in sample G. The resistance is deduced by fitting the low-bias part of IVs measured at different temperatures. A constant value of 650 \textOmega , corresponding to the source contact resistance, has been subtracted.}
        \label{fig:extrasG}
    \end{figure*}
    
\section{Additional measurements on sample G}
%Here the measurements with different sweep direction, raw data with multiple points, R(T). Perhaps three panels in two columns.
In this section we present additional experimental data from sample G. Figure~\ref{fig:extrasG}\textbf{a}  shows the raw data (namely, without subtraction of $-2.5$~mT in $B_z$ and 170~nA in current) used to generate the plot in Fig.~1\textbf{f} of the main text. For each value of $B_z$, ten IVs for positive bias and ten for negative bias where measured, whose critical current values are displayed in Fig.~\ref{fig:extrasG}\textbf{a}. A certain distribution in the switching current is visible. Each data point in Fig.1\textbf{f} of the main text is obtained averaging over these points. Without the averaging, the scatter in $I_c^{\pm}$ would produce a larger scatter in $Q$, owing to the fact that this latter quantity is obtained from the difference between $I_c^+$ and $|I_c^-|$. 

Figure~\ref{fig:extrasG}\textbf{b} shows measurements of the rectification efficiency $Q$ as a function of $B_z$ for different $B_z$-sweep directions. As for panel \textbf{a}, here we neither subtracted instrumental offsets, nor averaged the data. We notice that, within the data scatter, there is no apparent effect of the sweep direction.

Finally, Fig.~\ref{fig:extrasG}\textbf{c} shows a $R(T)$ curve as extracted from IV characteristics measured in a three-terminal configuration. Each data point is obtained by fitting the low bias part of the IV. The temperature independent value at low temperature ($R(0)\approx 650$~\textOmega) is the source contact resistance, which is unavoidable in a three-terminal measurement. From the $R(T)$ we deduce a $T_c=4.0$~K at the constriction. As for the other samples, this value is smaller than the typical $T_c$ for few-layer NbSe$_2$ ($>5$~K), owing to the disorder introduced in the constriction by etching.

 \begin{figure*}[tb]
        \centering
        \includegraphics[width=\textwidth]{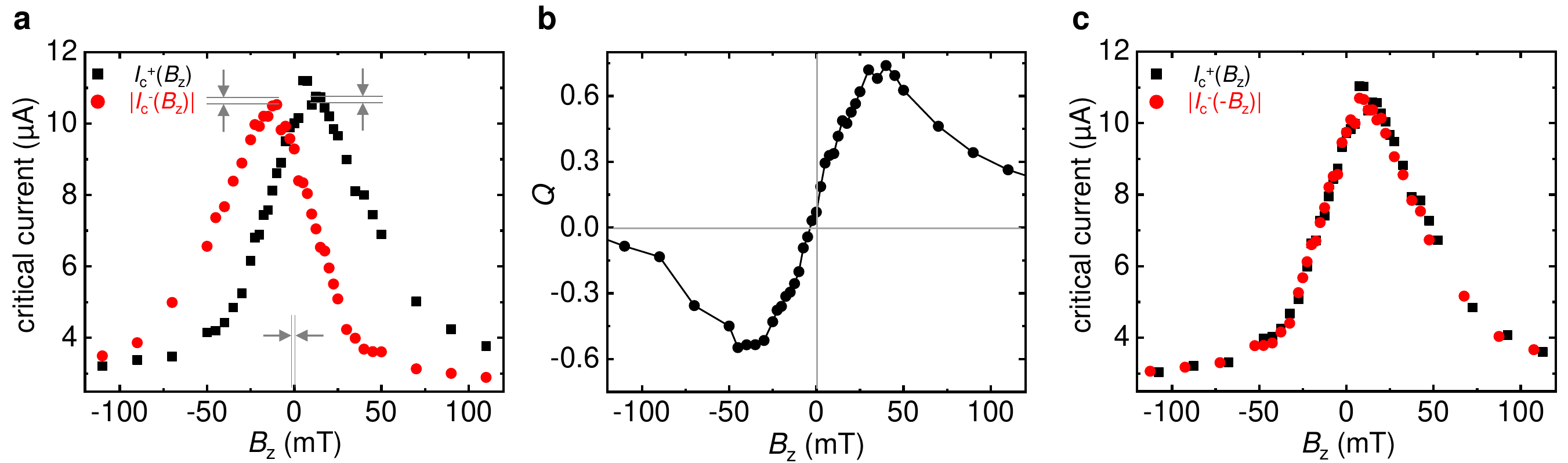}
        \caption{\textbf{Magnetic field and current offset in sample G.}  \textbf{a},~Positive and (absolute) negative critical current as a function of $B_z$ in sample G, displayed without applying offset subtraction, in contrast to Fig.~1\textbf{f} of the main text. Each point corresponds to the average of ten measurements. Grey bars and arrows indicate the magnitude of the offsets. As a result of the offset subtraction the $I_c^+(B_z)$ curve (black) is shifted downwards, while $|I_c^-(B_z)|$ is shifted upwards, and both are shifted in the horizontal positive direction.   \textbf{a},~Rectification efficiency $Q$ as a function of $B_z$ as deduced from panel \textbf{a}. In absence of offset subtraction the $Q(B_z)$ curve appears  shifted upwards by 0.1 compared to the graph in Fig.~1\textbf{g}. \textbf{a},~Plot of $I_c^+(B_z)$ and $|I_c^-(-B_z)|$ after applying an offset $B_z^{\text{off}}=-2.5$~mT and $I^{\text{off}}=170$~nA. This corresponds to Fig.~1{\textbf{f}} of the main text, with inversion of the abscissas for $|I_c^-|$.  
        }
        \label{fig:noff}
    \end{figure*}
    
\textit{Offset removal in sample G}.  
%In this section we shall discuss about the determination and removal of the offset for $B_z$ and for the supercurrent in Fig.~1\textbf{f,g} of the main text. 
In Fig.~\ref{fig:noff}\textbf{a} we show the same data as in Fig.~1\textbf{f} --namely, $I_c^+$ and $|I_c^-|$ for sample G-- but, in this case, without removing the field offset (-2.5~mT) and the current offset (170~nA). The graph is nearly indistinguishable from that in Fig.~1\textbf{f}, indicating that the offset has only a minor impact. 
In particular, the current offset is about 1.5\% of the maximum critical current. The resulting $Q(B_z)$ curve without offset removal is plotted in Fig.~\ref{fig:noff}\textbf{b}. As a result of the offset, the curve appears slightly shifted (both horizontally and vertically) compared to Fig.~1\textbf{g} of the main text. Note that the offset is relatively small, comparable with the width of the switching current distribution discussed above: its effect becomes visible only in the averaged $Q(B_z)$ curves, as the one shown in Fig.~1\textbf{g} or Fig.~\ref{fig:noff}\textbf{b}.  

A subtraction of $\delta B_z=-2.5$~mT and $\delta I_c=170$~nA from the nominal values of, respectively, $B_z$ and $I_c^{\pm}$\footnote{A subtraction of 170~nA to $I_c^-$ corresponds to an addition of the same amount to $|I_c^-|$.}, leads to a match of the $I_c^+(B_z)$ and $|I_c^-(-B_z)|$ curves, that is required by time-reversal symmetry. The overlapping curves are shown  Fig.~\ref{fig:noff}\textbf{c}. 

\section{Measurements on sample B}

  \begin{figure*}[tb]
        \centering
        \includegraphics[width=\textwidth]{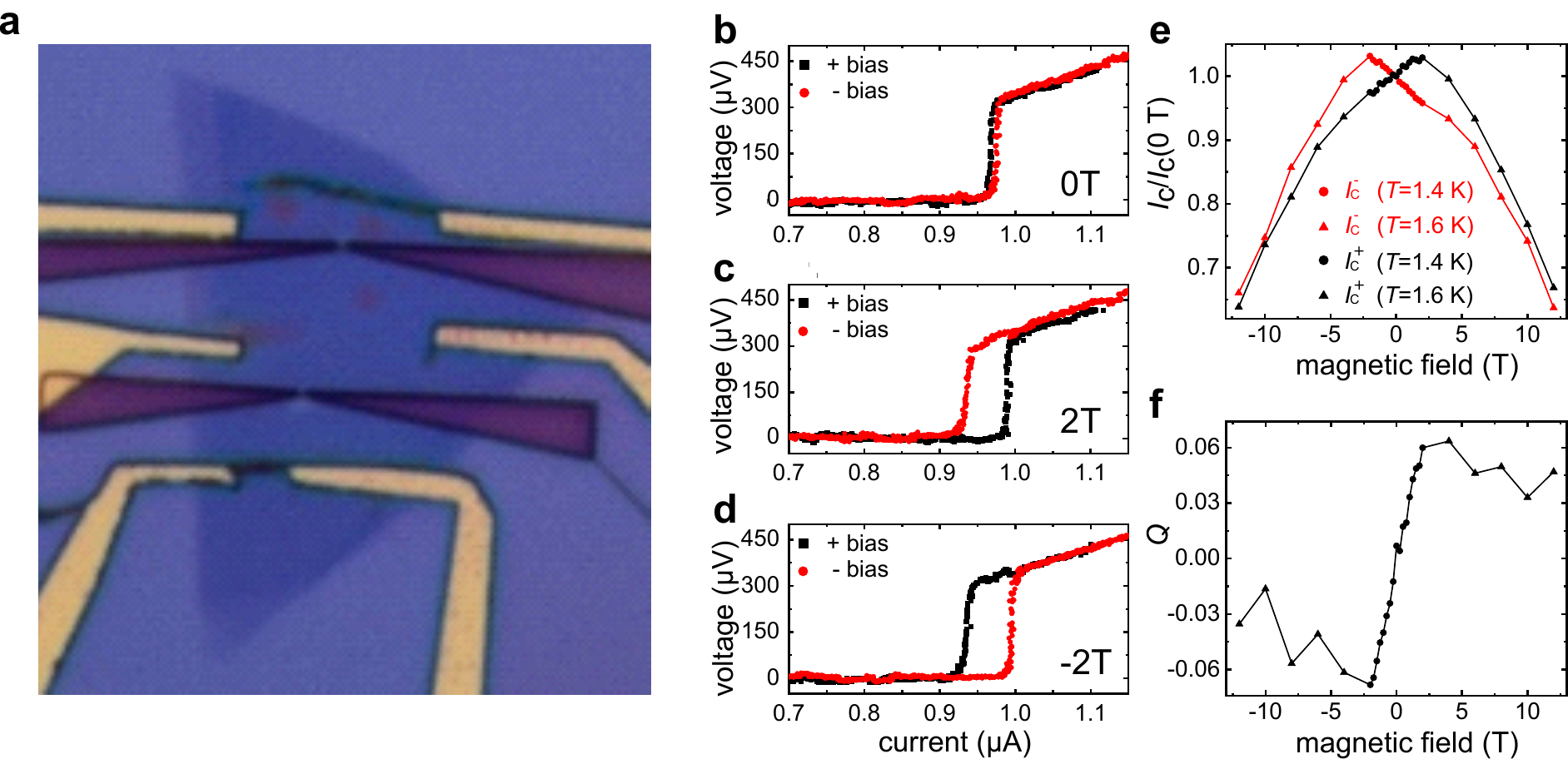}
        \caption{\textbf{Measurements on sample B.} \textbf{a}, Optical microscope picture of sample B.  \textbf{b},~Current-voltage characteristics (IVs) for opposite bias polarities (i.e., opposite current directions) measured in the absence of a magnetic field at $T=1.4$~K on sample B. The sweep direction is always from zero to finite bias. \textbf{c},~Similar measurements, but in the presence of a magnetic field $B=2$~T, mostly  directed in-plane but with an important additional $B_z$ component (see text). Notice the difference between the two critical currents. Importantly, the critical current for positive bias is larger than that in the absence of a field.  \textbf{d},~Same as in-panel \textbf{c}, but with opposite field orientation. The role of the two bias polarities is now swapped. \textbf{e}, Critical current for positive (black) and negative (red) bias as a function of the magnetic field. We combine two series of measurements at 1.4~K (circles) and 1.6~K (triangles). All values are normalized to that at zero field. \textbf{f}, Supercurrent rectification efficiency $Q\equiv 2(I_c^+-|I_c^-|)/(I_c^++|I_c^-|)$, plotted versus magnetic field.
        }
        \label{fig:sampleB}
    \end{figure*}

%Here basically what is in the old main text fig 1c--g.
In this section we discuss measurements on sample B. This sample (as well as samples D, E discussed below) was measured in the 1K cryostat with the magnetic field oriented nominally in-plane (and perpendicular to the current direction within  the constriction). Owing to a misalignment (of typically a couple of degrees) the field has also a small but decisive component out-of-plane which produces the supercurrent rectification.
Clearly, to obtain a $B_z$ field of tens of mT, $B_{ip}$ needs to be several teslas. On the other hand, owing to the characteristics of Ising superconductors, an in-plane field of few teslas has a little effect on the critical current.

Figure~\ref{fig:sampleB}\textbf{a} shows a magnified version without contour lines of the micrograph in Fig.~1\textbf{b} of the main text. Figures~\ref{fig:sampleB}\textbf{b}, \textbf{c} and \textbf{d} show IV-characteristics for an applied field of 0~T, 2~T and -2~T, respectively. This field magnitude corresponds roughly to the maximal rectification. Notice that the positive critical current for 2~T is clearly larger that that for 0~T: similar to sample G, sample B (as well as sample D and E, see below) shows an increase of the critical current with the field. for one polarity. Figure~\ref{fig:sampleB}\textbf{e} shows the dependence of the positive (black) and negative (red) critical current on the magnetic field. Each curve contains points from two different measurements performed with different resolution and at slightly different temperature (high resolution: $T=1.4$~K, circles; low resolution: $T=1.6$~K, triangles). For better comparison, the critical current are displayed normalized to the zero field value.
Finally Fig.~\ref{fig:sampleB}\textbf{f} shows the field dependence of the rectification efficiency $Q\equiv 2(I_c^+-|I_c^-|)/(I_c^++|I_c^-|)$. Panels \textbf{b-f} are the corresponding ones of panels \textbf{c-g} in the Fig.~1 of the main text. 

These measurements show that the qualitative behavior of sample B is similar to that of sample G. Owing to the unknown degree of misalignment, it is not possible to determine neither  $B_{\text{max},I_c}$ nor $B_{\text{max},Q}$. However, since by experience the misalignment is of the order of a couple of degrees, we estimate $B_z$ to be in the range of several tens of mT, which is also in line with sample G and sample F.

 \begin{figure*}[tb]
        \centering
        \includegraphics[width=\textwidth]{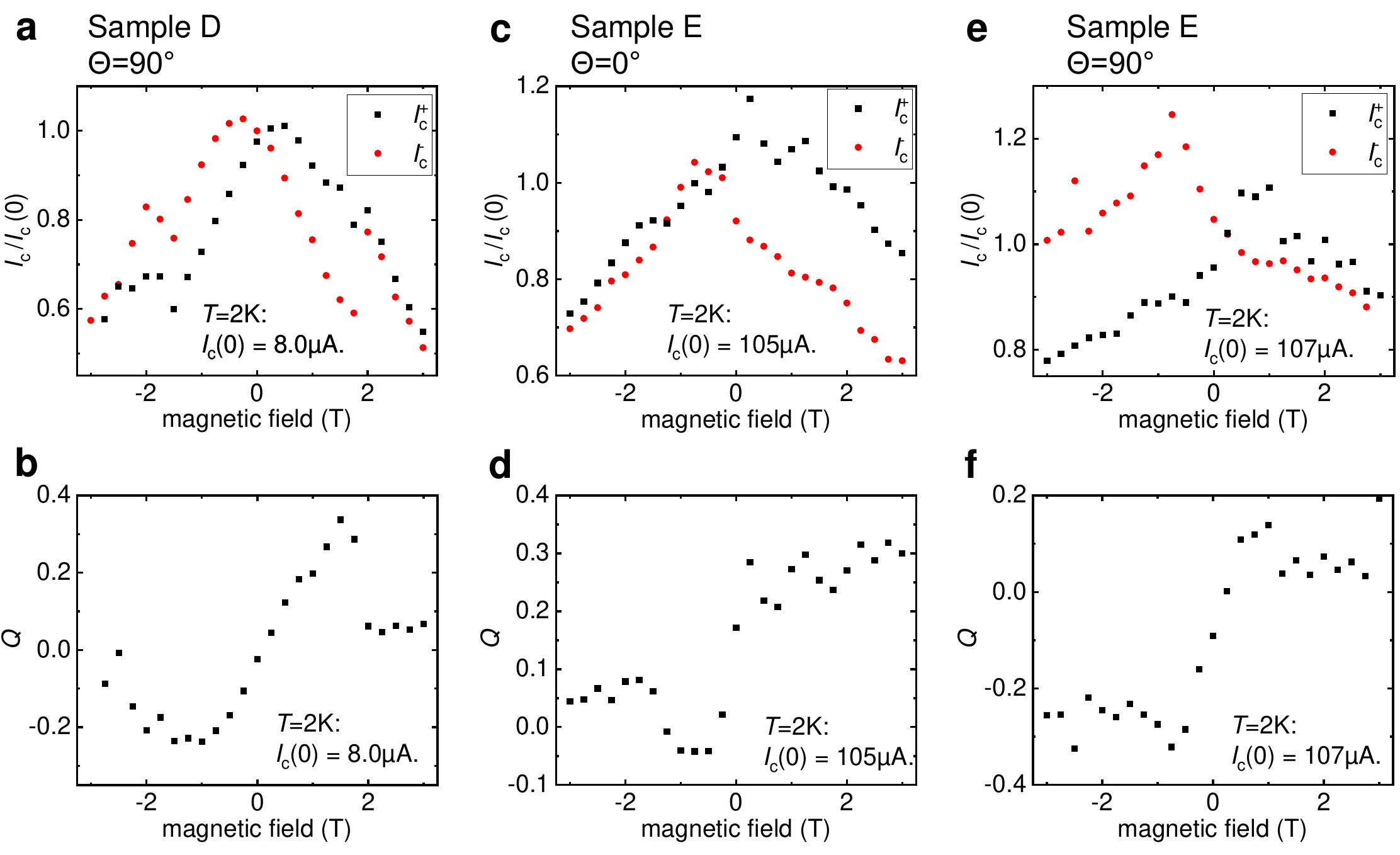}
        \caption{\textbf{Critical current and rectification efficiency in other samples.} \textbf{a,~c,~e},~Magnetic field  dependence of the critical current for positive ($I_c^+$) and negative ($I_c^-$) bias, measured in sample D (panel \textbf{a}), in sample E (panel \textbf{c} and \textbf{e}). The magnetic field is oriented mostly in-plane,  with an important (but unknown) out-of-plane component. The in-plane field component is perpendicular ($\theta=90^{\circ}$) to the current in \textbf{a} and \textbf{e}, while in panel~\textbf{c} it is parallel to the current. \textbf{b,~d,~f} Magnetic field  dependence of the rectification efficiency $Q$ measured in sample D (panel \textbf{b}) and in sample E (panel \textbf{d} and \textbf{f}). Each graph is obtained from the data in the panel immediately above. 
        }
        \label{fig:othersamples}
    \end{figure*}
   
  \section{Summary description of the other samples}
  
  In this section we briefly describe other samples investigated within this project which have not yet been discussed so far, namely, samples A, C, D, E. The letter used to label the samples indicated the sequence of their fabrication (sample A first, sample G at last).

  %Measurements on sample B and sample F are described in the main text. 
  %We also provide measurements of $\Delta I_c$ for samples D and E that were performed in a $^4$He cryostat equipped with a superconducting coil parallel to the sample surface, as for sample B (c.f.~Fig.~1 of the main text). 
  
  \textit{Sample A. } In this sample, measured in 4-terminal configuration, the central constriction stayed normal all the way down to the base temperature of out 1K cryostat. This was likely caused by an unintentionally more aggressive reactive ion etching step,  or by oxidation occurred after the lithographic fabrication of the constriction. 
  Therefore, this sample shall not be further considered here.

  \textit{Sample C. } This sample has a different geometry (1~\textmu m wide and 4 \textmu m-long channel) and a different substrate (ceramic substrate for solid ion gating experiments) compared to all the others. It does not show a well-defined  critical current: instead, it displays a series of phase slip lines, which some of the authors have studied in plain few-layer NbSe$_2$ crystals~\cite{Paradiso_2DMat_2019}. Owing to the peculiarities of this sample, we shall not consider it further.
  %Therefore the critical currents are not signaling the pair-breaking limit,
  %but only the nucleation threshold for the phase slip line nucleation. Despite the differences (the sample was designed for different investigations) we have nevertheless investigate a possible difference in the critical currents for the nucleation of a phase slip lines.    
  
  \textit{Sample D. } This sample (nominally identical to A, B, E, F). The plot of the two critical currents $I_c^+(B)$ and $I_c^-(B)$ is similar to that in Fig.~1\textbf{f} of the main text. Similarly the plot of $Q(B)$ mirrors that in Fig.~1\textbf{g}. 
  Critical current and supercurrent rectification efficiency measurements, reported in Fig.~\ref{fig:othersamples}\textbf{a}, show a behavior similar to that observed in sample B, c.f.~Fig.~1 of the main text. Also in this device the absolute value of the critical current increases with $|B|$, reaching a maximum for $|B|\simeq 0.4$~T. This value is smaller than that for sample B (about 2~T), most likely owing to a larger field misalignment in sample D. This sample shows the second-largest rectification efficiency: for $B=1.5$~T, $Q$ is as large as 33\%, see Fig.~\ref{fig:othersamples}\textbf{b}. 
  
  \textit{Sample E. } The critical current of this device is very large. The $I_c$ value, well above 100~\textmu A, is close to the critical current for an entire flake and it is clearly incompatible with a width of few hundreds nanometers which is that of the constriction. The most likely interpretation is that, due to some undetected problems during the reactive ion etching step, the etching depth did not exceed that of the top hBN layer, leaving the NbSe$_2$ flake unaffected. As a consequence, the direction of the current is not well-defined as for the other samples. Despite this, we do observe a clear supercurrent diode effect, see Fig.~\ref{fig:othersamples}\textbf{c-f}.
  From SHG measurements we deduce that the supercurrent direction within the constriction is nearly  parallel (2$^{\circ}$ misalignment) to the armchair direction of the lattice, see next section. 
  
  We measured   sample E in two cool-downs. In the first cool-down the applied field was mostly directed in-plane and parallel to the nominal constriction direction. In the second, the sample was rotated by $90^{\circ}$ so that the field was still mostly directed in-plane, but perpendicular to the nominal constriction direction. The lack of a well-defined constriction makes it difficult to disentangle the effect of the different field components. Possibly because of that, the $I^{\pm}_c(B)$ and the $Q(B)$ plots appear asymmetric and distorted. It is important to remark that a supercurrent rectification is nevertheless  observed, confirming the robustness of the diode effect in NbSe$_2$.

    \clearpage 
    \newpage
    
    \begin{figure*}[b!]
        \centering
        \includegraphics[width=\textwidth]{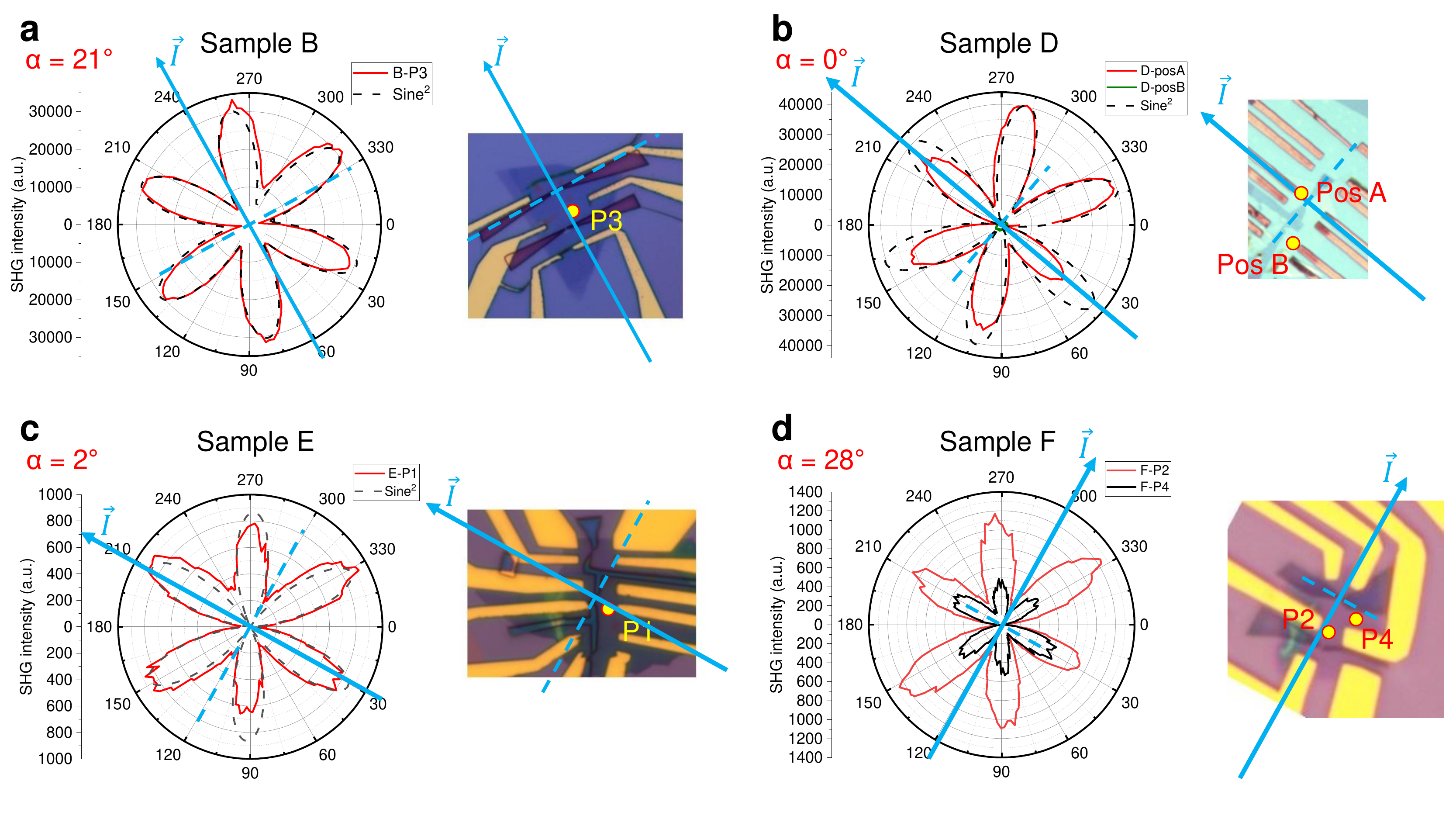}
        \caption{\textbf{Second harmonic generation measurements.} \textbf{a},~Polar plot of the second harmonic generation (SHG, red curve) signal measured on sample B in the position indicated by the yellow dot (position P3) in the optical micrograph displayed on the right side. The dashed line in the graph shows a $\sin^2$ fit. The supercurrent direction (vector $\vec{I}$) is indicated on both sides by a light blue arrow. On the micrograph, the relevant constriction is located where the perpendicular dashed line crosses the arrow. The angle $\alpha$  between $\vec{I}$ and the armchair direction is indicated in red on the top left. \textbf{b},~The same for sample D. The red curve (pos A) corresponds to the 3-layer NbSe$_2$ area where the constriction is located. The weak signal shown by the green curve (pos B) corresponds to another area with even number of layers. \textbf{c},~The same for sample E. \textbf{d},~The same for sample F. The position P4 is closer to the constriction but both signals refer to the same layer number.
        }
        \label{fig:shg}
    \end{figure*}

\section*{Second harmonic generation measurements}

To determine the armchair direction and layer number parity of the NbSe$_2$ crystal, we measure the co-polarized SHG intensity as a function of relative angle between laser polarization and crystal orientation~\cite{Xi2015Ising,MalardSHG13,LiSHG13,Lin2019} A Ti:sapphire laser (Spectra-Physics, Mai Tai XF, 80~MHz repetition rate, 80~fs pulse duration) at 800~nm was focused onto the NbSe$_2$ samples placed in the vacuum chamber using a microscope objective (Olympus, LUCPLFLN 40$\times$) with numerical aperture of 0.6. The reflected SHG signal at 400~nm was collected with the same objective, filtered by a 680~nm short-pass filter (Semrock, FF01-680SP), dispersed in a spectrometer (Princeton Instruments, Acton SP2300) with a 150 grooves/mm grating, and detected by a CCD camera (Princeton Instruments, PIXIS 100). A linear polarizer was placed in front of the spectrometer to ensure acquisition of the signal parallel to the laser polarization. A 50:50 non-polarizing beam splitter was used to separate the incident path and signal detection path. An achromatic half-wave plate was placed between the beam splitter and the objective to change the relative angle between the crystal orientation and the laser polarization. The half-wave plate was rotated in 1$^{\circ}$ increments from 0$^{\circ}$ to 180$^{\circ}$ using a stepper motor, and the SHG intensity was recorded after each step. In general, a laser power of 1~mW and a single exposure time of 1~s were used for each measurement. 

The polar plots in Fig.~\ref{fig:shg} show the typical six-fold symmetry pattern representing the three-fold symmetry of the transition metal dichalcogenide crystals. The maximum (minimum) intensity direction corresponds to the armchair (zigzag) direction of the crystal. The light blue arrow indicates the axis of the constriction, i.e.,  the direction of the supercurrent in the transport measurements. The arrow is reproduced on the optical micrograph on the right side of each panel: there, the perpendicular  dashed line crosses the arrow exactly at the position where the constriction is located.  The yellow dot indicates the position where the corresponding SHG measurement indicated in the graph was performed. In the figure, panel \textbf{a}, \textbf{b}, \textbf{c}, and \textbf{d} refer to sample B, D, E, and F, respectively. The angle $\alpha$ between supercurrent and armchair direction is indicated in red on the top left of each panel. Only for sample G, the angle dependence is not reported: owing to its even number $N$ of layers, the SHG signal is very small, except at the very edges. This sample is also very homogeneous, therefore it does not feature terraces with odd $N$ which could allow us to perform an angle-resolved SHG measurements with a discernible signal. SHG measurements on this sample were nevertheless precious, since they confirmed its parity and demonstrated unambiguously that the supercurrent  diode effect can be observed in crystals with both even and odd $N$.

From the \textit{combination} of optical microscopy and SHG measurements, we deduce that samples B, D, E, F, G consist of 3, 3, 5, 3, and 2 layers, respectively. In fact, optical microscope pictures do not make it always possible an univocal determination of the layer number: a typical optical micrograph is typically compatible with two consecutive values of $N$.The parity determination provided by SHG eliminates the residual uncertainty in $N$.

Concerning the direction of the supercurrent with respect to the underlying lattice, in sample D and in sample E the vector $\vec{I}$ is almost perfectly aligned parallel to the armchair direction while in sample F is nearly parallel to the zig-zag direction. Finally, in sample B the orientation is intermediate, but closer to zig-zag. 

% \textit{Sample B. } This sample has been discussed in the main text, Fig.~1. From second harmonic generation measurements (SHG) we deduce that the supercurrent direction within the constriction is displaced by 21$^{\circ}$ from the armchair direction of the lattice, see next section. 

%\textit{Sample F. } This sample has been discussed in the main text, Fig.~2. From SHG measurements we deduce that the supercurrent direction within the constriction is displaced by 28$^{\circ}$ from the armchair direction of the lattice, i.e., it is nearly parallel to the zig-zag direction, see next section. 

We notice that the different magnitude of the supercurrent rectification factor among the samples does not seem to be correlated to the angle $\alpha$. The same applies to the other peculiar phenomenon observed in our experiment, namely, the increase of the critical current with the magnetic field. This effect produces an opposite horizontal shift of the $\Lambda$-shaped $I_c^+(B)$ and $I_c^-(B)$ curves, see Fig.~1\textbf{f} of the main text and Fig.~\ref{fig:othersamples}\textbf{a}-\textbf{c} here. The only sample where this effect is negligible (see Fig.~2\textbf{d} of the main text) is F, where $\vec{I}$ is oriented approximately along the zig-zag direction. On the other hand, in sample B the effect is particularly pronounced, despite an angle $\alpha$ relatively close to that for sample F.

    \clearpage 
    \newpage

\section{The negligible role of vortices}
\begin{figure}[tb]
        \centering
        \includegraphics[width=\columnwidth]{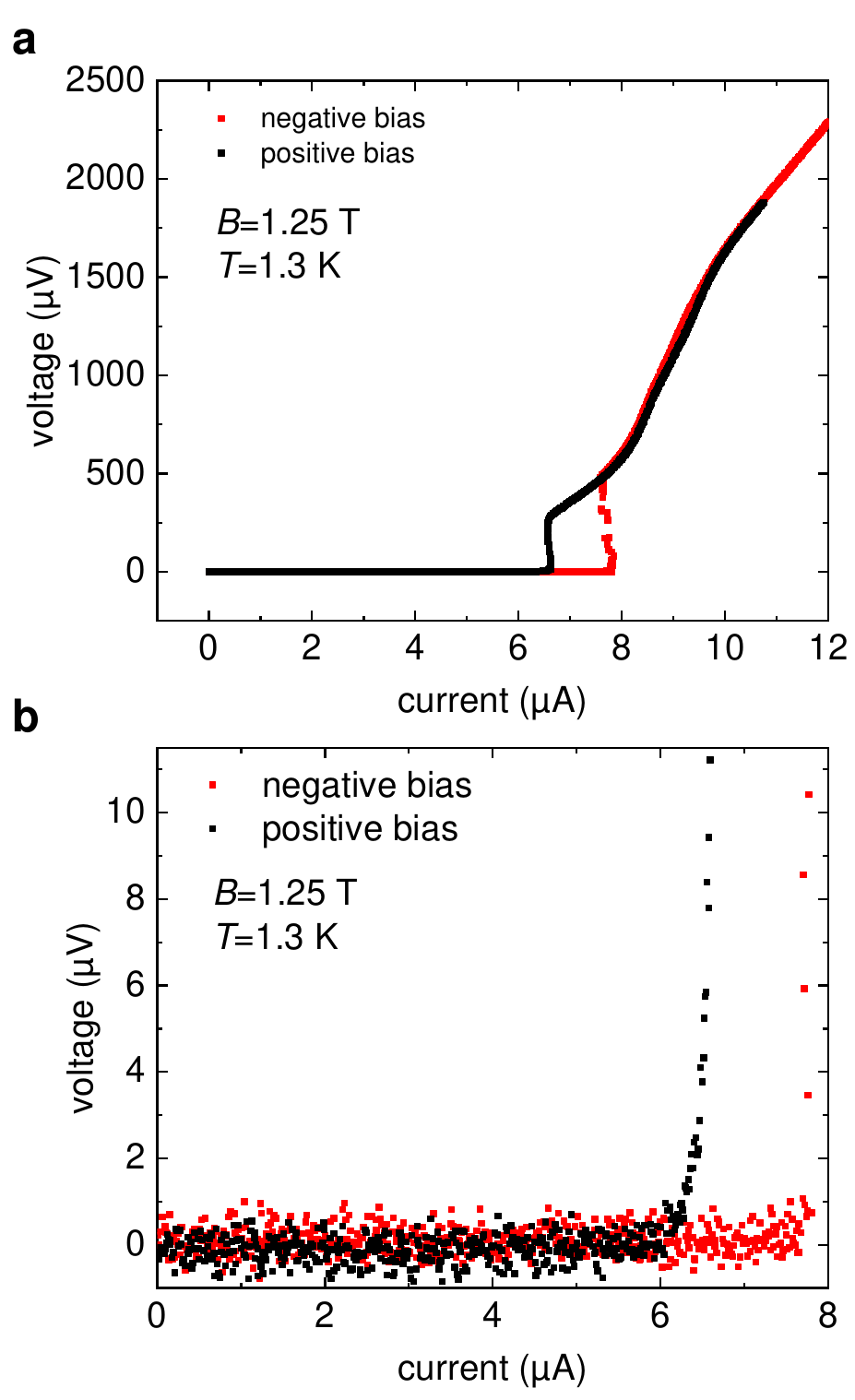}
        \caption{\textbf{Emergence of dissipation in sample D.} \textbf{a},~Current-voltage characteristics measured on sample D. The black (red) symbols refer to the positive (negative) bias polarity. \textbf{b},~Zoom-in on a voltage scale of $\approx 10$ \textmu V. It is evident that dissipation emerges abruptly from the noise floor at the critical current value, without showing the typical smooth foot produced by vortex creep.
        }
        \label{fig:exIVsampleD}
    \end{figure}

In 2D Rashba superconductors the supercurrent diode effect is driven by in-plane magnetic field, as shown by experiments~\cite{Ando2020,baumgartner2021diode} and predicted by theory~\cite{he2021arx}. Therefore, the presence of vortices can be, in principle,   eliminated by careful compensation of the out-of-plane component. In contrast, in NbSe$_2$ (and, more generally, in materials with valley-Zeeman spin-orbit interaction) it is precisely the out-of-plane field component that drives the supercurrent nonreciprocity, leading to the unavoidable presence of vortices. These might conceivably  introduce spurious nonreciprocity, e.g. in the presence of an asymmetric  barrier for entering/leaving the samples at the two edges of the constriction. However, the experimental evidence indicates that the presence of vortices does not play a significant role, owing to the following arguments.
\begin{enumerate}
    \item We do not observe a measurable dissipation in the IVs until we reach the critical current, and this remains true up to fields at least as large as $B_{\text{max},Q}$. As an example, in Fig.~\ref{fig:exIVsampleD} we show IV characteristics for sample D (where 4-terminal IV-characteristics are available). In this case, as described above, the field is applied mainly in-plane, with an important out-of-plane component (of the order of several tens of mT). Notice that there is no measurable foot within a voltage scale of 500~nV: at the critical current, the voltage emerges abruptly from the noise floor. 
    \item If the diode effect were due to the asymmetry in the barrier for vortices  entering/leaving the sample, the rectification would be extremely temperature dependent (following the temperature dependence of the penetration depth $\lambda$ and thus of the vortex barrier) while approaching $T_c$. Instead, the $T$-dependence in sample D and F is even nonmonotonic, and in general it does not follow the typical $T$-dependence of $\lambda$.
    \item If the vortex barrier played a significant role, one would expect a significant vortex trapping within the constriction, leading to hysteresis in the field sweep. Instead, as shown in Fig.~\ref{fig:extrasG}\textbf{b}, the $B_z$-sweep direction is irrelevant. 
    \item If the diode effect were due to vortices, the in-plane $\mathbf{B}_{ip}$ field would not make the effect of $B_z$ asymmetric, as we observed in the experiments. \textit{A fortiori}, the $\mathbf{B}_{ip}$ sign and orientation would be totally irrelevant, in contrast to results shown in Fig.~2 of the main text.
    %\item The observed nonmonotonic $Q(B_z)$ dependence (nearly linear at moderate field, then rapidly  suppressed for  $B_z>B_{\text{max},Q}$) is hard to explain through a mechanism based only on vortices since  both $B_{\text{max},I_c}$ and $B_{\text{max},Q}$ are much lower than $B_{c2}$ for NbSe$_2$ ($\approx 3$~T), therefore lower than any relevant threshold field related to vortex physics.
    %\item In a $250$~nm$ \times 250$~nm constriction, a magnetic flux quantum $\Phi_0$ corresponds to $B_z \approx 32$~mT. But we observe a clear and reproducible diode effect at field much smaller than that (e.g., the linear regime for $Q(B_z)$ is typically within $|B_z|<10$~mT in sample G).
\end{enumerate}

     \begin{figure*}[tb]
        \centering
        \includegraphics[width=\textwidth]{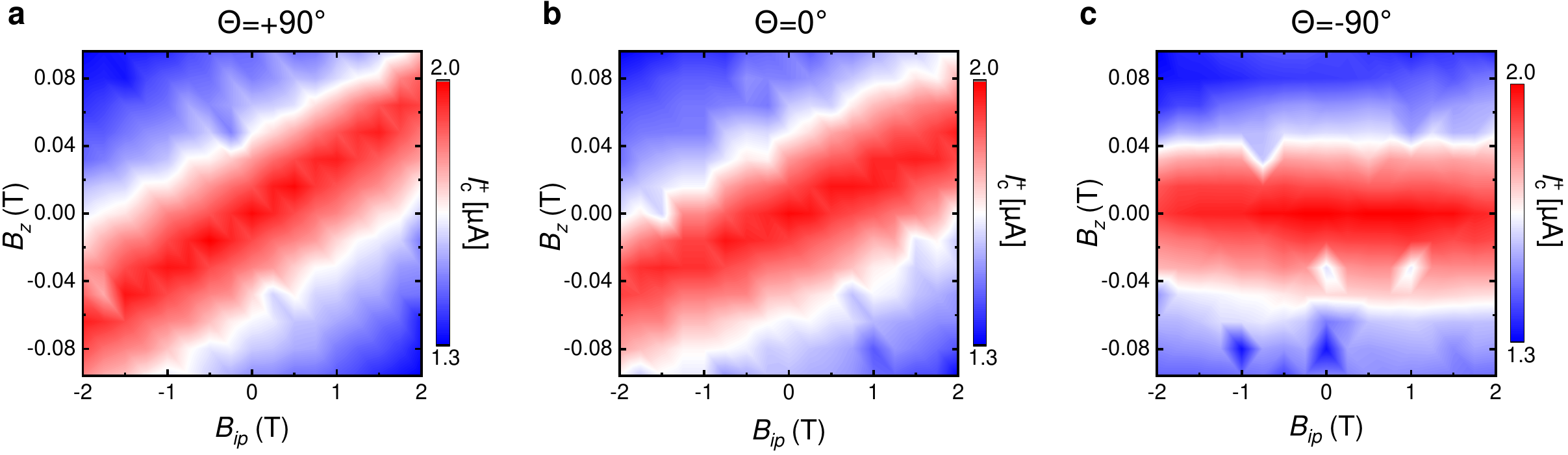}
        \caption{\textbf{Critical current as a function of in- and out-of-plane field for different $\theta$ values.} \textbf{a},~Color plot of the positive critical current $I_c^+$ as a function of the \textit{nominally applied}  in-plane ($B_{ip}$) and out-of-plane ($B_z$) field, measured for an angle $\theta=90^{\circ}$ between current and $\vec{B}_{ip}$, c.f.~Fig.~2\textbf{a} of the main text. The maximum of each $I_c^+(B_z)$ curve (vertical cut of the color plot) occurs at a field $B_{z,\text{max}}$, which depends linearly on $B_{ip}$, indicating a misalignment as described in the text. \textbf{b}~The same, measured for $\theta=0^{\circ}$,  c.f.~Fig.~2\textbf{b} of the main text. \textbf{c}~The same, measured for $\theta=-90^{\circ}$,  c.f.~Fig.~2\textbf{c} of the main text.
        }
        \label{fig:mis}
    \end{figure*}  
    
\section*{Magnetic field misalignment}

As discussed in the main text, the alignment of the magnetic fields with respect to the sample surface is crucial for the interpretation of the data. In fact, the perturbing effect of $\vec{B}_{ip}$ is noticeable for fields of the order of teslas. On the other hand, a couple of degrees of field misalignment would produce a spurious out-of-plane field of the order of tens of milliteslas, more than sufficient to induce a sizeable diode effect. 

The sample misalignment is due to two factors. First, the sample rotation axis might be not perfectly perpendicular to $\vec{B}_{ip}$. Second, the sample surface might not be perfectly perpendicular to the rotation axis. Owing to both misalignment effects (each of the order of about 1 or 2 degrees) the angle between $\vec{B}_{ip}$ and the sample surface varies as a function of $\theta$ with $2\pi$-periodicity. As a consequence, measurements with $B_z$ at fixed $B_{ip}$ will be affected by an offset proportional to  $B_{ip}$, with constant of proportionality being a $2\pi$-periodic function of $\theta$.

The determination of such an offset (and thus of the misalignment) can be easily achieved by measuring the critical current (either $I_c^+$ or $I_c^-$) as a function of the nominal $B_z$ and $B_{ip}$ fields. For an Ising superconductor the application of a few teslas in-plane field has little impact on the critical current, while out-of-plane fields of even a few tens of milliteslas have a noticeable effect. Therefore, if we plot, as in Fig.~\ref{fig:mis}, $I_c^+(B_z,B_{ip})$, the $B_z(B_{ip})$ offset can be determined from the local maximum. For a given $B_{ip}$ the maximum of $I_c^+(B_z)$ must occur when the \textit{effective} $B_z$ field is zero. This condition allows us to find, for that value of $\theta$, the additional  $B_z$ introduced by $B_{ip}$. 
We deduce that $B_z=B_{z,\text{nom}}+\kappa B_{ip}$, where $B_z$ is the effective out-of-plane field (used in the main text, see Fig.~2a-c), $B_{z,\text{nom}}$ is the nominally applied out-of-plane field, and $\kappa$ is a coefficient such that $\kappa(90^{\circ})=0.0343$, $\kappa(0^{\circ})=0.0218$, and $\kappa(-90^{\circ})=-0.00256$. Approximately, from measurements at other intermediate angles $\theta$, we found that $\kappa(\theta)\approx 0.015+0.019\sin (\theta-0.30)$, with $\theta$ expressed in radians.

     \begin{figure*}[tb]
        \centering
        \includegraphics[width=0.9\textwidth]{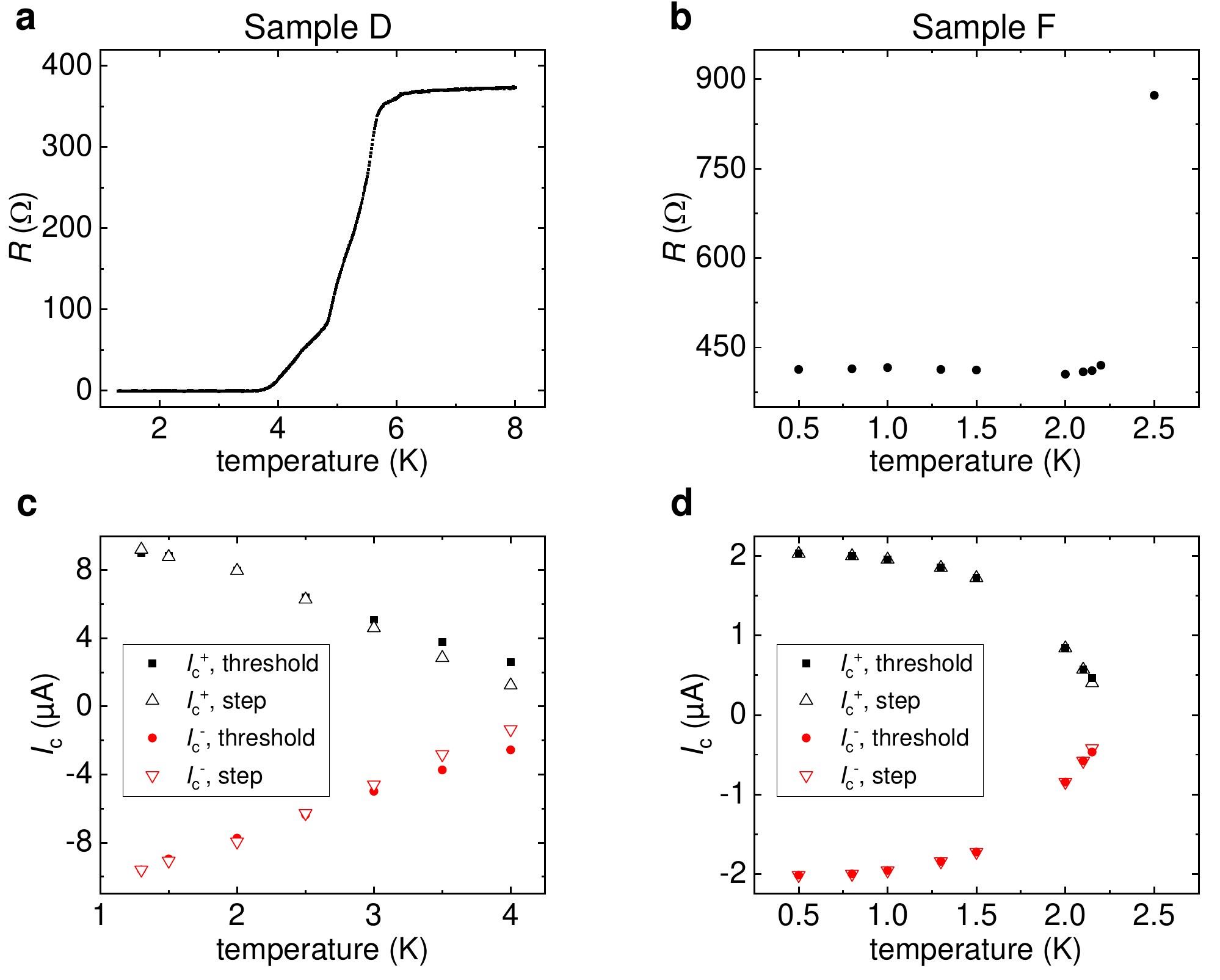}
        \caption{\textbf{$R(T)$ and $I_c^{\pm}(T)$ for sample D and F.} \textbf{a},~Resistance versus temperature measured in a 4-terminal configuration on sample D. The first emergence of a detectable resistance above the noise occurs at $T=3.68$~K.  \textbf{b},~Resistance versus temperature measured in a 3-terminal configuration on sample F. The low temperature resistance (413~\textOmega) corresponds to the contact resistance. The first emergence of a detectable resistance above the noise occurs at $T=2.2$~K. \textbf{c},~$I_c^{+}$ (black) and $I_c^{-}$ (red) versus temperature for  sample D. Full symbols refer to critical current values deduced from the threshold condition $V=V_{thr}\equiv 100$~\textmu eV. Empty symbols refer to critical current values deduced by extrapolating the steepest portion of the IV to the abscissa axis. The latter points converge to $T_c=4.3$~K. 
          \textbf{d},~The same as in panel \textbf{c}, but for sample F. From the convergence of the empty symbols to the abscissa axis we deduce $T_c=2.25$~K. 
        }
        \label{fig:RT}
    \end{figure*}   
\section*{Resistance and critical current versus temperature}

The graphs in Fig.~\ref{fig:RT}\textbf{a},\textbf{b} show the zero-bias resistance versus temperature for the constriction in sample D and F, respectively. Measurements on sample D were performed in 4-terminal, by combining an AC (lock-in amplifier) and a DC (digital-to-analog converter) voltage source, and by measuring the output current with a current amplifier and the voltage drop with a differential voltage amplifier. $R(T)$ curves are obtained from the zero-bias differential resistance measured with the lock-in amplifier, while the critical current is obtained from the DC IVs.
Measurements on sample F were performed in 3-terminal configuration (since one of the 4 contacts did not work), using only DC excitation. $R(T)$ data are obtained from the low-bias slope of the IVs, upon subtraction of the constant contact resistance of 413~\textOmega.
The temperature for which the resistance emerges from the noise floor of the deep superconducting state is 3.68~K in sample D and 2.2~K in sample F.

The graph in Fig.~\ref{fig:RT}\textbf{c} shows the temperature dependence of the positive and negative critical current. We stress that, as throughout this work, the IVs from which we determine $I_c$ are always swept from zero bias to finite (either positive or negative) bias, in order to eliminate any possible heating effect. The full symbols in  Fig.~\ref{fig:RT}\textbf{c}  refer to critical current values deduced from the threshold condition $V=V_{thres}\equiv 100$~\textmu eV. The empty symbols refer to critical current values deduced by extrapolating on the abscissas the steepest portion (corresponding to the critical bias step) of the IV characteristics. Note that, except for very close to $T_c$, the critical current criterion is immaterial, since the IVs are discontinuous at $I_c$. Close to $T_c$, instead, the $V=V_{thres}$ criterion tends to overestimate the critical current, since a finite resistance (and thus a finite voltage) might appear at low bias due to uncompensated vortices that get depinned, or to vortex-antivortex depairing close to the Berezinskii-Kosterlitz-Thouless (BKT) transition. On the other hand, the alternative criterion  (extrapolation on the current axis of the steepest portion of the IV) is difficult to apply near $T_c$ since the step itself becomes highly smeared. In Fig.~\ref{fig:RT}\textbf{c} we   plotted only points where the IVs display a sufficiently  well-defined voltage step. From the convergence of the open symbols on the $T$-axis ($I^{\pm}_c(T=T_c)=0$ condition), we deduce a critical temperature $T_c=4.3$~K. This value is slightly higher than that ($T=3.68$~K) corresponding to the emergence of a finite zero-bias resistance. This discrepancy is probably due to the aforementioned dissipative processes occurring below the mean field $T_c$.

Figure~\ref{fig:RT}\textbf{d} is the analogue of Fig.~\ref{fig:RT}\textbf{c} for sample F. In this case from the convergence of the empty symbols onto the $T$-axis we deduce $T_c=2.25$~K, which is very close to the temperature value ($T=2.2$~K) where a finite resistance emerges.

In the main text, we have used as $T_c$ values the ones deduced from the convergence of the $I_c^{\pm}(T)$ curves onto the temperature axis, i.e., the $I_c^{\pm}(T=T_c)=0$ criterion. Therefore we assigned $T_c=4.3$~K for sample D and $T_c=2.25$~K for sample F. 

\section*{Outliers in Fig.~2f of the main text}
\begin{figure}[tb]
        \centering
        \includegraphics[width=\columnwidth]{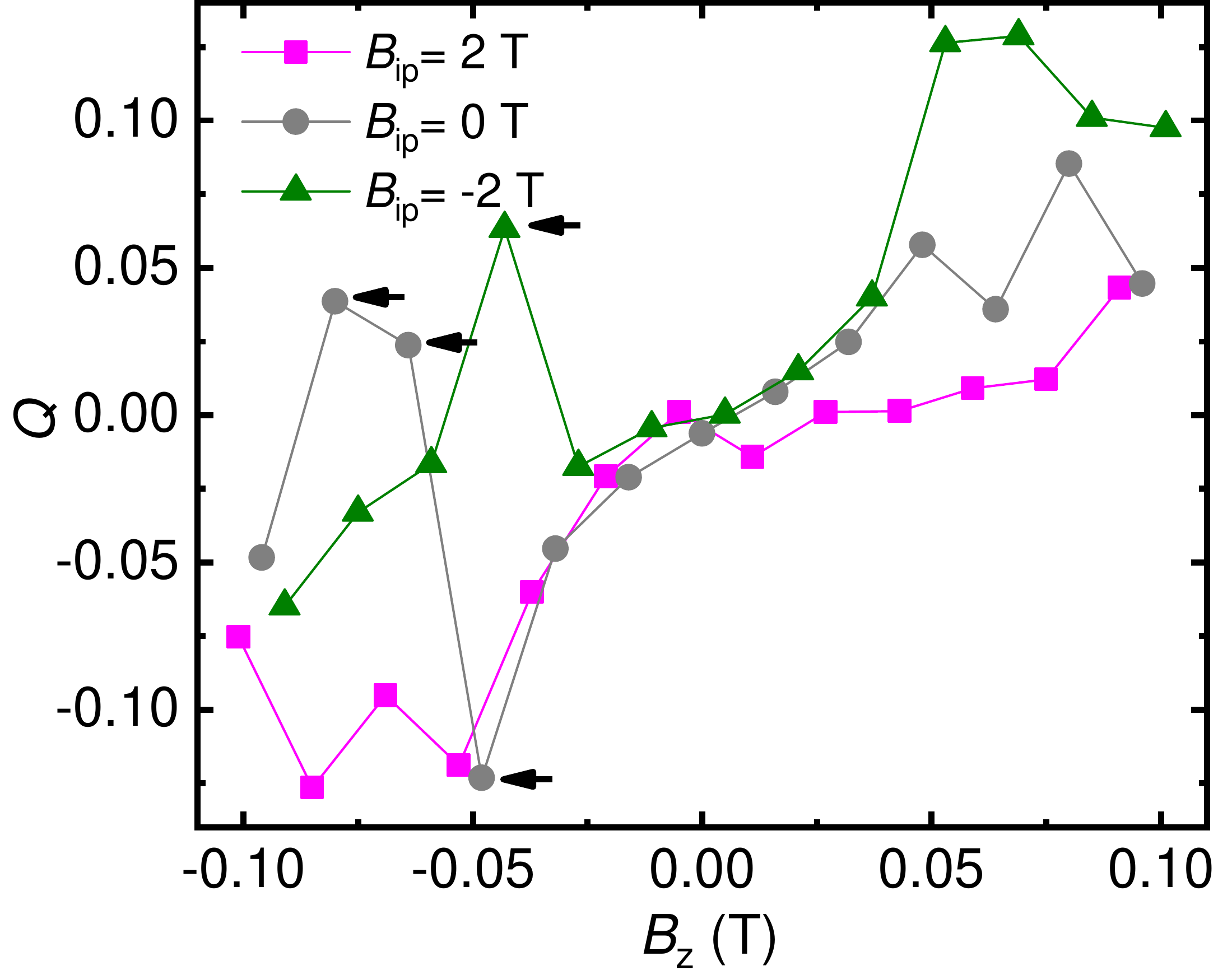}
        \caption{\textbf{Figure 2f without outlier removal} The graph shows the data in Fig.~2\textbf{f} of the main text, without the outlier removal described therein. The arrows indicate the four points which were substituted in the main part of the article, see text. 
        }
        \label{fig:outl}
    \end{figure}

In  Fig.~2\textbf{f} of the main text four points (three data points, $B_z = -48$, $-64$ and $-80$~mT for the curve $B_{ip}=0$~T;  one data point $B_z = -43$~mT for the curve $B_{ip}=-2$~T) were strong outliers. To reduce scatter we substituted the outliers with the corresponding points measured for the adjacent $B_{ip}$ value (i.e., $B_{ip}=-0.25$~T instead of 0~T,  $B_{ip}=-1.75$~T instead of -2~T). Figure~\ref{fig:outl} shows the original data. The outliers are indicated by arrows. Outliers are also visible in the color plots in Fig.~2\textbf{a-c}. They might may conceivably originate from temperature instabilities during the measurements.

\end{document}